\documentstyle[prb,multicol,aps,epsf]{revtex}

\newcommand{\tJ}{$t$-$J$\ }

\newcommand{\fluxvar}{\langle\Phi^2\rangle}

\begin{document}

\draft
\title{Bosons, gauge fields, and high $T_c$ cuprates}
\author{Don H. Kim, Derek K. K. Lee, and Patrick A. Lee}
\address{Department of Physics, Massachussetts Institute of Technology,
Cambridge, MA, 02139}
\date{\today}
\maketitle
\begin{abstract}
  A simple model of a degenerate two-dimensional Bose liquid
  interacting with a fluctuating gauge field is investigated as a
  possible candidate to describe the charge degree of freedom in the
  normal state of the cuprate superconductors.  We show that the
  fluctuating gauge field efficiently destroys superfluidity even in
  the Bose degenerate regime. We discuss the nature of the resulting
  normal state in terms of the geometric properties of the
  imaginary-time paths of the bosons. We will also present numerical
  results on the transport properties and the density correlations in
  the system. We find a transport scattering rate of $\hbar/\tau_{\rm
  tr}\sim 2k_{\rm B}T$, consistent with the experiments on the
  cuprates in the normal state. We also find that the density
  correlations of our model resemble the charge correlations of the
  $t$-$J$ model.
\end{abstract}

\pacs{PACS numbers: 74.20.Mn, 67.40.Db}

\begin{multicols}{2}
\narrowtext

We study the low-temperature behavior of repulsive bosons in a
spatially fluctuating gauge field in two dimensions. This is motivated
by the gauge theories of the $t$-$J$ model for the cuprate
superconductors, where low-energy charge excitations are described by
bosonic degrees of freedom. The internal gauge field of this model
suppresses superfluidity in the Bose liquid, even below the Bose
degeneracy temperature when there is significant exchange among the
bosons. We can study the imaginary-time trajectories of the bosons in
the path-integral representation of this model. We see that the boson
world-lines retrace themselves in the presence of strong gauge
fluctuations, giving rise to interesting dynamics in this degenerate
but metallic Bose liquid.

We have studied this metallic state using quantum Monte Carlo
techniques. We find that this model does indeed capture some of the
long-wavelength charge properties which are common to the cuprate
superconductors. This includes a linear temperature dependence of the
transport scattering rate $1/\tau_{\rm tr}$, as deduced from a
Drude-like optical conductivity from our model. This is consistent with
experimental data on the cuprate superconductors near optimal doping.
We also find that the density excitations in our model are
qualitatively similar to those in the full $t$-$J$ model, by comparing
our results with diagonalization results in the literature. A brief
account of this work has already appeared \cite{derek}.

\section{motivation}

The normal metallic state of the superconducting cuprates displays
many non-Fermi-liquid properties. For instance, the in-plane
resistivity of La$_{2-x}$Sr$_x$CuO$_4$ has a power-law temperature
dependence of the form $\rho \propto T^\alpha$ where $\alpha$
increases from 1 to 1.5 with increasing hole doping \cite{batlogg}. In
particular, near optimal doping, the resistivity is linear in
temperature up to 1000K. This linear-$T$ dependence is found in many
of the cuprate superconductors with similar values of $d\rho/dT$
(1.2$\mu\Omega$cm/K $\pm$ 20\%) \cite{ong}. This should be contrasted
with the quadratic temperature dependence of Fermi-liquid theory.
Similarly, the transport relaxation rate appears to be universal among
optimally-doped compounds: $1/\tau_{\rm tr} \simeq 2k_{\rm B}T$ (from
a two-component-model analysis of the optical conductivity in
YBCO\cite{orenstein}, LSCO\cite{gaoLaSr}, Bi2212\cite{romeroBi},
Bi2201\cite{romeroBi}). Transport in a magnetic field is also
anomalous. The Hall coefficient indicates the existence of hole-like
carriers in the doping range where superconductivity occurs. The Hall
coefficient $R_{\rm H}$ increases with decreasing temperature, but it
remains smaller than the classical value of $1/n_hec$ for a hole
density of $n_h$ for a wide range of temperatures down to the
superconducting transition. These compounds also have a small positive
magnetoresistance with a temperature dependence\cite{harris} different from
conventional theory using $\tau_{\rm tr}$.

The transport properties of these compounds appear to have common
features in spite of considerable differences in the transition
temperature and spin fluctuation properties among these
compounds. This indicates that a common mechanism is responsible for
the scattering of charge carriers in these materials.  One might hope
that this scattering mechanism can be understood in terms of a
low-energy theory with a minimum number of microscopic parameters. In
this paper, we study a Bose liquid in a fluctuating gauge field as a
possible candidate for such an effective theory.

The anomalous transport behavior, together with other unusual features
such as temperature-dependent magnetic susceptibility and non-Korringa
behavior of the nuclear magnetic relaxation time, leads to the
conclusion that the metallic state of the cuprates cannot be
described in a simple Fermi-liquid scenario. It has been postulated that
``spin-charge separation'' is responsible for these
anomalies\cite{pwa87}. For instance, such a scenario might reconcile
the apparent low density and hole-like character of the charge
carriers with the observation of a large, electron-like Fermi surface
in photoemission. 
Numerical studies of the $t$-$J$ model, which is believed to be a
low-energy model of the cuprates, also provide some support for
spin-charge separation \cite{putikka,horsch,eder}, such as different
energy scales for the spin and charge excitations, and the suppression
of 2$k_F$-scattering in the charge spectrum.

A model of spin-charge separation is a gauge theory where neutral
spin-half fermions (``spinons'') and charge-$e$ bosons (``holons'')
interact {\em via\/} an internal U(1) gauge field \cite{ioffelarkin,PN}.
Physically, the transverse part of the gauge field is related to
``spin chirality'' fluctuations \cite{PN}. In this picture, the charge
properties of the system should be dominated by the behavior of the
holons. We will study the holon subsystem in this paper, treating the
spinon subsystem simply as a medium through which the gauge field
propagates. To be more precise, we study a model of bosons with
on-site repulsion in the presence of a spatially fluctuating magnetic
field with short-range correlations. The repulsive interaction is
necessary for the stability of the system, which means that one cannot
treat this problem perturbatively starting from an ideal Bose
gas. Previous studies
\cite{wheathong,wheatscho,menkeRG,menkepath,ioffekal} have implicitly
studied the non-degenerate regime of low density or high temperature,
whereas the regime relevant to the cuprates is the degenerate regime
where the thermal deBroglie wavelength of the bosons is greater than
the mean particle spacing.  A concern from earlier studies of the
gauge model is that degenerate bosons would have strong diamagnetic
response to the internal gauge field and hence effectively
Bose-condense at a relatively high temperature ($k_{\rm B}T_{BE} \sim
4\pi n_h t \sim 1000$K). This would in fact restore Fermi-liquid
behavior to the system. We shall show here that gauge fluctuations
suppress this diamagnetic response and the bosons remain normal
without strong diamagnetism at all finite temperatures. Furthermore,
our numerical results indicate that the resistivity of this Bose
metallic phase has a linear temperature dependence which is consistent
with experiments.

Besides the possible relevance to the transport in the cuprate
superconductors, the model we consider is of intrinsic theoretical
interest. The model is a Bose version of the problem of a quantum
particle in a random magnetic flux, which has received considerable
attention in recent years. 
It is also related to frustrated spin systems and vortex glasses.
However, since we deal exclusively with annealed averaging in this
paper (see later), we cannot draw any direct conclusions about these
problems with quenched disorder.

The rest of the paper is organized as follows.  In section II, we
review the connection between the gauge theory of the $t$-$J$ model
and our boson model. In section III, we discuss the path-integral
formalism which provides a convenient framework to visualize physical
processes in terms of the imaginary-time paths of the bosons.  In
section IV, we look at the effects of the gauge field on the
world-line geometry of the bosons. We will see that the partition
function of the system is dominated by self-retracing world-line
configurations. We will also argue that superfluidity is destroyed by
the fluctuating gauge field, giving rise to a degenerate Bose
metal. In the subsequent sections, we present the results of a quantum
Monte Carlo study of this metallic phase. We will discuss the transport
properties and the density correlations in this boson model.

\section{A Boson Gauge Model}

In this section, we provide the motivation for studying an effective
boson model from the gauge theory of the $t$-$J$
model, which describes the motion of vacancies in a doped Mott
insulator:
\begin{equation}
{\cal H}= -t_0 \sum_{\langle ij\rangle \sigma}
(c_{i\sigma}^{\dagger}c_{j\sigma} + {\rm  h.c.})
+ J\sum_{\langle ij\rangle} {\bf S}_i\cdot{\bf S}_j 
\end{equation}
with the constraint of no double occupancy. Experimentally, $J\simeq$
1500K and $t_0/J\simeq 3$. 

The constraint of no double occupancy allows us to write the
creation of a physical hole in terms of the creation of a
charged hard-core boson (holon) and the annihilation of a spin-half
fermion (spinon): $c_{i\sigma}=f_{i\sigma}b_i^{\dagger}$.  In terms of
these slave bosons and fermions, the Hamiltonian of the $t$-$J$ model
can be written as:
\begin{eqnarray}
{\cal H}&=& -t_0 \sum_{\langle ij\rangle \sigma} 
(f_{i\sigma}^{\dagger}b_ib_j^{\dagger}f_{j\sigma} + {\rm h.c.})
+ J\sum_{\langle ij\rangle} {\bf S}_i\cdot{\bf S}_j   \nonumber\\
&+&\sum_i a_{0i}(f_{i\sigma}^{\dagger}f_{i\sigma} + b_i^{\dagger}b_i -1)
\label{tJ}
\end{eqnarray}
where ${\bf S}_i= f_{i\alpha}^{\dagger}\sigma_{\alpha\beta} f_{i\beta}$,
The $a_{0i}$-field is a Lagrange multiplier enforcing the local
occupancy constraint, and acts as a fluctuating scalar potential for
the spinons and holons.

Among the mean field theories proposed to decouple the quartic terms
in Eq.\ (\ref{tJ}), a candidate for the normal state near optimal
doping is the the uniform resonating-valence-bond (RVB) ansatz:
$\sum_{\sigma}\langle f_{i\sigma}^{\dagger}f_{j\sigma}\rangle
=\xi e^{ia_{ij}}$. This incorporates
short-range antiferromagnetic correlations without any long-range
N\'eel order.
The Lagrangian of this RVB phase can be written as:
\begin{eqnarray}
{\cal L}&=& \sum_{i,\sigma} f_{i\sigma}^{\ast}
(\partial_{\tau}-\mu_{\rm F}+a_{0i})f_{i\sigma}+
\sum_i b_i^{\ast}(\partial_{\tau}-\mu_{\rm B}+a_{0i})b_i
\nonumber \\ 
&&\quad - \frac{J}{2}\xi \sum_{\langle ij\rangle} (e^{ia_{ij}}
f_{i\sigma}^{\ast}f_{j\sigma} + {\rm h.c.}) \nonumber\\
&&\quad - t_0\xi \sum_{\langle
ij\rangle} (e^{ia_{ij}} b_i^{\ast}b_j + {\rm h.c.})
\label{rvb}
\end{eqnarray}

The vector potential $a_{ij}$ arises from the fluctuations in the
phase of the RVB order parameter. Longitudinal fluctuations of the
gauge field $a_{ij}$ do not affect the Lagrangian due to an internal
U(1) gauge symmetry:
\begin{eqnarray}
f_i&\longrightarrow& f_i e^{i\theta_i} \nonumber\\
b_i&\longrightarrow& b_i e^{i\theta_i} \nonumber\\
a_{ij}&\longrightarrow& a_{ij}-\theta_i+\theta_j 
\end{eqnarray}
We will therefore work in a fixed gauge, such as the Coulomb gauge,
and consider only the fluctuations in the transverse part of the gauge
field $a_{ij}$. In other words, we will consider only
fluctuations in the internal magnetic and electric fields which are
gauge-invariant quantities.

Since we are interested in the charge degrees of freedom, we wish to
consider an effective theory with bosons only, and regard the spinon
fluid as a medium through which the gauge field propagates.  The gauge
field has no dynamics {\em in vacuo}. The response of the spinon fluid
to the gauge field is responsible for the dynamics of the gauge field
as seen by the holons. More specifically, we can obtain the Gaussian
fluctuations of the $a$-fields by treating the spinon response in the
random-phase approximation. The effective gauge-field propagator is:
\begin{eqnarray}
{\cal S}_{\rm G}&=& \frac{1}{2\beta L^2}\sum_{{\bf k},\omega_n}  
\Pi^{00}({\bf k},\omega_n)a_0^{\ast}({\bf k},\omega_n) a_0({\bf k},\omega_n) 
\nonumber \\
&+& \frac{1}{2\beta L^2}\sum_{{\bf k},\omega_n}\Pi^{\perp}({\bf k},\omega_n)
a_{\perp}^{\ast}({\bf k},\omega_n) a_{\perp}({\bf k},\omega_n),
\end{eqnarray}
where $\beta=1/T$, $\omega_n=2\pi nT$, $L$ is the linear size of the
system, and $a_{\perp}$ is the transverse part of the gauge field.
(We use units where distance is measured in terms of the lattice
spacing and $k_{\rm B}=\hbar=e=1$.)  Here, for small $k$ and
$\omega_n$, $\Pi^{00} \simeq \rho_{\rm F}$, the spinon density of
states at the Fermi level. This describes the Thomas-Fermi screening
of internal electric fields by the fermions. The effective interaction
mediated by the screened $a_0$-field is a repulsion between the bosons
(of range $\propto\rho_{\rm F}^{-1/2}$), consistent with the original
hard-core requirement for the bosons. We will model this with an
on-site repulsion energy, $U$. On the other hand, the magnetic fields
due to fluctuations in $a_{ij}$ are not effectively screened out by
the fermions\cite{reizer}. The gauge-field fluctuations as experienced
by the holons are therefore strong. More specifically, the Gaussian
fluctuations have the correlation function $D({\bf
k},\omega_n)=\langle a_{\perp}^{\ast}({\bf k},\omega_n)a_{\perp}({\bf
k},\omega_n) \rangle$, given (in the continuum limit) by:
\begin{equation}
D({\bf k},\omega_n)= \frac{1}{\Pi_{\perp}({\bf k},\omega_n)}=
\frac{1}{\gamma |\omega_n|/k + \chi k^2}
\label{gaugeprop}
\end{equation}
where $\chi$ is the orbital susceptibility of the spinon fluid and
$\gamma$ is a Landau damping coefficient. These gauge-field
fluctuations cause profuse forward scattering of the bosons. We
believe that this is the dominant scattering mechanism in this
problem. Since it is overdamped at long wavelengths with a relaxation
rate which diverges as $1/k^3$, we will ignore the slow relaxation and
work in a ``quasistatic'' limit for the gauge fields:
\begin{equation}
D({\bf k},\omega_n)\rightarrow D({\bf k},\omega_n=0)\,\delta_{n,0}=
\delta_{n,0}/\chi k^2.
\label{quasistat}
\end{equation}
(On a square lattice, $k^2$ is replaced by $1-(\cos k_x - \cos
k_y)/2$.)  This quasistatic approximation is justified when the gauge
field relaxes on a time scale longer than $1/T$. Since the typical
scattering wavevector of interest is the inverse deBroglie wavelength
of the bosons, the relevant relaxation time scales as $1/k^3\sim
1/T^{3/2}$. One might therefore expect\cite{ikwieg} that this
approximation is valid at a sufficiently low temperature. 


One might object that arguments above are based on a weak-coupling
theory of the response of the spinons to the gauge fields. However, we
believe that the essential features remain correct in general, namely
a separation of times scales between the relaxation of the gauge
fields and the boson dynamics, as well as the magnitude of the gauge
fluctuations being controlled by the spinon diamagnetic susceptibility
$\chi$.

The gauge-field correlator (\ref{quasistat}) corresponds to a
spatially uncorrelated flux distribution with the correlation
function:
\begin{equation}
\langle \Phi_{\bf r}\Phi_{\bf r'} \rangle = 
\frac{T}{\chi}\,\delta_{{\bf r},{\bf r'}}
\label{fluxtemp}
\end{equation}
where $\Phi_{\bf r}= (\Phi_0/2\pi) \sum_\Box a_{ij}$ (oriented sum
around the links of plaquette ${\bf r}$) is the flux through plaquette
${\bf r}$. ($\Phi_0 = hc/e$ is the flux quantum.) Since we are
treating the thermodynamics for the gauge field classically, we have a
thermal factor of $T$ in Eq.\ (\ref{fluxtemp}) for the flux variance
$\langle \Phi^2 \rangle$. Given that the fermion orbital
susceptibility is roughly constant at low temperatures, we might
expect the flux variance to have a linear temperature dependence.
However, a lattice calculation by Hlubina {\em et al.}\cite{hlubina}
has indicated that the Gaussian fluctuations are sufficiently strong
that the flux through a plaquette is of the order of the flux quantum
$\Phi_0$: $\fluxvar^{1/2}\geq 0.5\Phi_0$ down to a temperature of
$0.4J$. Since the experimental superconducting $T_c$ is of the order
of $0.1J$, we expect that this regime of strong random flux is
relevant to the normal state of the cuprates until one approaches
the superconducting transition. In this regime, the precise value of
$\fluxvar$ should not affect the behavior of the bosons, and we will
focus on a {\em large and temperature-independent\/} flux variance when
we study the transport and correlation functions of our boson system.

Another factor leading to the reduction of the flux variance at low
temperature is one that has not been discussed so far, namely that the
magnitude of the gauge field should also be affected by the
diamagnetic response of the holons as well as the spinons, {\em i.e.},
$\fluxvar=T/(\chi_{\rm spinon}(T)+\chi_{\rm holon}(T))$. The holon
contribution dominates near an instability to Bose condensation where
$\chi_{\rm holon}$ diverges and the bosons develop a Meissner response
to expel the gauge field from the system altogether. However, we will
see in this paper that Bose condensation and the holon diamagnetism
are strongly suppressed even below the boson degeneracy temperature.
Therefore, in a wide range of temperatures above the superconducting
$T_c$, we are justified in neglecting this feedback effect of the
holons on the magnitude of the gauge field fluctuations.

We can now define more precisely the effective model which we study in
the rest of the paper. It is a model of lattice bosons interacting
with a quasistatic gauge field, described by effective action
$S=S_{\rm B}+ S_{\rm G}$:
\begin{eqnarray}
S_{\rm B} &=& \int_0^{\beta}\!
        \left( \sum_i  b_i^{\ast} (\partial_{\tau}-\mu_{\rm B}) b_i
                - H_{\rm B}(\tau) \right) d\tau   \nonumber \\
S_{\rm G} &=& \frac{1}{2\beta L^2}\sum_{{\bf k}}
        D^{-1}({\bf k},0) |a_{\perp}({\bf k},0)|^2
\label{modelgauge}\\
          &=&\sum_{\bf r}\frac{\Phi_{\bf r}^2}{2\langle \Phi^2 \rangle}  
\label{modelflux}
\end{eqnarray}
with the boson Hamiltonian
\begin{equation}
  H_{\rm B} = -t \sum_{\langle ij\rangle} ( e^{ia_{ij}}
  b_i^{\dagger}b_j + {\rm h.c.} ) + \frac{U}{2} \sum_i n_i (n_i -1),
\label{hamilton}
\end{equation}
where $t = t_0 \xi \sim t_0$, $L$ is the linear size in units of
lattice spacing and $U \gg t$. Note that, on performing the average
over the gauge field, we average over static configurations only, {\em
i.e.}, ${\bf a}({\bf k},\omega_n\neq 0)=0$.

We cannot say that we have rigorously derived above effective action
from the slave-boson mean field theory of the $t$-$J$ model. Many
approximations have been introduced to obtain this simple model with few
adjustable parameters. For example, we have neglected the temperature
dependence of the RVB order parameter $\xi$ and also the gauge-field
correlations of higher order\cite{gauss}. We take the point of view
that we are studying a ``minimal'' low-energy theory which hopefully
captures many of the generic features of more complicated models.

\section{Path Integral Representation}

\begin{figure}
\epsfxsize=\columnwidth\epsfbox{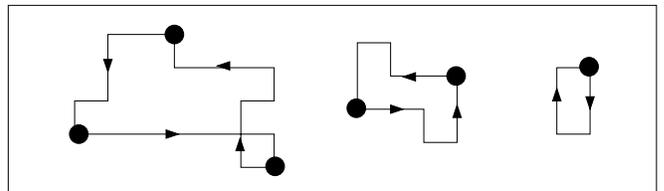}
\caption{A schematic configuration for 6 bosons after projecting the
imaginary-time paths onto the $xy$-plane. There are a total of 3
cycles: 1 cycle of one particle, 1 cycle of two particles, and 1 cycle
of three particles. Solid circles denote particle positions at
$\tau=0$ and $\beta$.}
\label{cycles}
\end{figure}

It is convenient to study our boson model in a first-quantized
formulation. The partition function $Z$
for a system with $N$ bosons in the canonical ensemble can be written
in terms of a Feynman path integral\cite{feynman} over the boson
trajectories $\{{\bf x}_\alpha(\tau)\}$ ($\alpha=1,\ldots,N$):
\begin{eqnarray}
Z &=& \frac{1}{N!}\sum_P  
\!\int^{{\bf x}(0)=P({\bf x}(\beta))}\!\!\!\!\!\!\!\!{\cal D}
[{\bf x}_1,\ldots,{\bf x}_N]\times\nonumber\\
&&\int \!\! {\cal D}{\bf a} \,\delta(\nabla\cdot{\bf a})
 e^{-S_{\rm G}({\bf a})-i\sum_{\alpha} 
\int_0^{\beta}{\bf a}\cdot {\bf\dot x}_{\alpha} d\tau-
S_{\rm B}^0(\{{\bf x}\})}
\label{pathpart}
\end{eqnarray}
where $S_{\rm B}^0$ is the action for bosons in the absence of
magnetic fields:
\begin{equation}
S_{\rm B}^0= \int_0^{\beta}\!\!d\tau \,
\left(\sum_i b^*_i \partial_{\tau} b_i - H_{\rm B}^0\right)
\end{equation}
where $H_{\rm B}^0$ is given by (\ref{hamilton}) with $a_{ij}=0$. In
this section, we will discuss the model in the continuum limit for
notational convenience. In the continuum, one would have:
\begin{equation}
S_{\rm B}^0=
  \int_0^{\beta}\!\!d\tau \,\left[ \sum_{\alpha=1}^{N} \frac{m}{2}{\bf\dot
  x}_{\alpha}^2 + \sum_{\alpha > \gamma} U \delta({\bf
    x}_{\alpha}(\tau)-{\bf x}_{\gamma}(\tau)) \right].
\end{equation}
Particle identity is taken into account by performing the path
integral over all trajectories where the set of final boson
coordinates at $\{{\bf x}_1(\beta),\ldots,{\bf x}_N(\beta)\}$ is some
permutation of the initial boson coordinates $\{{\bf x}_1(0),\ldots,
{\bf x}_N(0)\}$. Any such permutations can be broken down to
cycles. Each cycle forms a closed loop when the imaginary-time
trajectories (world lines) of a many-boson configuration are projected
onto real space. At high temperatures, cycles of length 1 dominate the
partition function and the system is in a non-degenerate classical
regime. At temperatures below the degeneracy temperature of the
bosons, particles can travel large distances in the imaginary time,
forming many ring exchanges (see Fig.~\ref{cycles}).

In this formulation, we may integrate out the Gaussian fluctuations of
the gauge field in (\ref{pathpart}). Thus, we arrive at a boson-only
effective theory which we study numerically in this work. The
system is described by the partition function $Z=\int\!{\cal
D}{\bf x} e^{-S_{\rm eff}}$ where the effective action is given by:
\begin{equation}
  S_{\rm eff} = S_{\rm B}^0+S_2
 \label{effective}
\end{equation}
with
\begin{equation}
        S_2 = {\textstyle\frac{1}{2}} \sum_{\alpha\alpha'}
                \int_0^{\beta}\!\!\int_0^{\beta}
                \!\!{\tilde  D}({\bf x}_\alpha(\tau)-{\bf
                x}_{\alpha'}(\tau'))\, 
                {\bf\dot x}_{\alpha}\!\cdot\!{\bf\dot x}_{\alpha'}
                \,d\tau \,d\tau'.
\label{currint}
\end{equation}
where ${\tilde D}({\bf x})= (1/\beta L^2)\sum_{{\bf k}\neq 0} D ({\bf
k},0) e^{-i{\bf k}\cdot{\bf x}}$. Note that the ${\bf k}=0$ contribution
has been excluded in the sum over ${\bf k}$, corresponding to
a gauge choice where the ${\bf k}=0$ part of ${\bf a}$ is zero. This
is one way to fix the remaining degree of gauge freedom which is not
determined by the condition of $\nabla\cdot{\bf a} = 0$. If we
consider a system with periodic boundary conditions in space, another
scheme would be to fix the line integral of the gauge field around a
specified path which wraps around the boundary. However, the latter
scheme is inconvenient for our purposes because it breaks
translational invariance explicitly.

The current interaction $D({\bf x})$ mediated by the
gauge field is logarithmic at large distances, and is attractive
between opposite currents. Due to the quasistatic nature of the gauge
fields, the interaction is also infinitely retarded in time. We will
see in the next section that this encourages world lines to retrace
themselves, with important consequences for the boson dynamics.

Before proceeding to discuss the physical consequences of the current
interaction $S_2$, some remarks about our averaging procedure for the
gauge fields are in order. We have performed an ``annealed'' average
over the gauge fields, rather than a ``quenched'' average. Annealed
averaging is necessary in our case because our gauge field ${\bf a}$
is an internal thermodynamic variable. Formally, we evaluate
observables $\langle{\cal O}\rangle$ as:
\begin{equation}
\frac{1}{Z}\,\int\!\!{\cal D}{\bf x}\, {\cal O} e^{-S_{\rm eff}}
=
\frac{\int\!{\cal D}{\bf x}{\cal D}{\bf a}\,P[{\bf a}]\,{\cal O}\, 
e^{-S_{\rm B}^0-i\int{\bf a}\cdot d{\bf x}}}
{\int\!{\cal D}{\bf x}{\cal D}{\bf a}\,P[{\bf a}]
\,e^{-S_{\rm B}^0-i\int{\bf a}\cdot d{\bf x}}}
\end{equation}
where $P[{\bf a}] = {\cal N}^{-1}\delta(\nabla\cdot{\bf a}) e^{ -
S_{\rm G}[{\bf a}]}$ is the probability distribution for the gauge
field, and ${\cal N}$ is a suitable normalization factor.  This is
different from quenched averaging which would be appropriate if we
dealt with a system with frozen impurities, such as a vortex glass.
Quenched averaging requires the evaluation of:
\begin{equation}
\int\!\!{\cal D}{\bf a} \,P[{\bf a}]\,
\left[\frac{\int\!{\cal D}{\bf x}\,{\cal O}\, 
e^{-S_{\rm B}^0-i\int{\bf a}\cdot d{\bf x}}}
{\int{\cal D}{\bf x}\,e^{-S_{\rm B}^0-i\int{\bf a}\cdot d{\bf x}}}\right].
\end{equation}
%
The differences between quenched and annealed averaging from the point
of view of perturbation (diagrammatic) theory has been addressed
elsewhere\cite{menkepath,hertz}.

From the point of view of the path integral Monte Carlo method, our
ability to perform the annealed averaging means that we would not have
to perform extensive averages over different frozen realizations of
the random flux. Moreover, note that the effective action
(\ref{effective}) is manifestly real, and so we avoid the sign problem
which occurs numerically when performing a quenched average over the
gauge fields.
We have studied boson densities between $n_b=1/4$ and $1/6$. We choose
an on-site interaction strength $U\ge 4t$. We follow the Monte Carlo
methods of Ceperley and Pollock\cite{ceppol} and Trivedi\cite{triv}.
Each Monte Carlo step involves the reconstruction of the world lines,
$\{{\bf x}_\alpha(\tau)\}$, for all $N$ particles using the ideal
boson propagator in a short interval in imaginary time. The on-site
interaction and the current interaction $S_2$ are taken into account
using Metropolis tests. To ensure quantum exchange, we may insist that
each accepted configuration differs from the previous one by a pair
exchange. This can be incorporated, without loss of detailed balance,
as a Metropolis test. We refer readers to the original
references\cite{ceppol,triv} for further details. (In evaluating the
gauge field contribution $S_2$, we have also made use of a geometrical
interpretation of $S_2$ which we discuss in the next section.) In the
discretization of the imaginary time, we have used a small
$\Delta\tau=\beta/M\leq 0.1/t$, so as to minimize the systematic error
and to allow the reliable use of maximum entropy techniques to perform
analytic continuation on our imaginary-time data to obtain the
dynamical quantities of interest. This sets the lowest accessible
temperature to $T\sim 0.1t$ for lattice sizes considered here. For
studies on dynamic response to be discussed later, we have restricted
ourselves to lattices of sizes up to $6\times 6$, due to the need to
obtain imaginary time correlation functions to a high accuracy. For
the calculation of static properties, we have studied lattices as
large as $10\times 10$.

To summarize, we have obtained an effective theory of bosons with
current interactions which are long-ranged in space and time. This
model can be studied using path integral Monte Carlo methods. In the
next section, we will discuss how these interactions affect the
geometry of the boson world lines and hence the physical properties of the
system.

\section{Effect of Gauge fields on world line geometry}
\subsection{``Brinkman-Rice bosons''}

In this section, we will discuss how the current interaction $S_2$
mediated by the gauge field affects world-line geometry. On the
infinite plane, there is a simple geometrical interpretation of this
interaction in terms of the winding numbers of the boson world
lines. The winding number $w_{\bf r}$ around a plaquette ${\bf r}$ is
the number of times the imaginary-time world lines of all the bosons
wind around the plaquette. Consider the partition function before
averaging over the gauge field. The effect of the gauge field enters
the partition function as the phase factor $\exp[-i\sum_\alpha
\int {\bf a}\cdot d{\bf x}_\alpha]$ in Eq.\ (\ref{pathpart}) over the
gauge field. This phase factor can be written in terms of $w_{\bf r}$:
$\sum_\alpha \int {\bf a} \cdot d{\bf x}_\alpha =\sum_{\bf r} w_{\bf
r} \Phi_{\bf r}$. We can now perform the average directly over the
Gaussian flux distribution (\ref{modelflux}), instead of the gauge
field distribution (\ref{modelgauge}). We will be working with
periodic boundary conditions ({\em i.e.}, on a torus). This will be
well-defined if we impose a constraint of zero total flux through the
system. On averaging, the phase factor becomes:
\begin{eqnarray}
\lefteqn{\int\!d\lambda\!\int\!\prod_{\bf r} d\Phi_{\bf r} \,\, 
e^{-\sum_{\bf r} \frac{\Phi_{\bf
r}^2}{2\fluxvar} - \frac{2\pi i}{\Phi_0}\sum_{\bf r} w_{\bf r}\Phi_{\bf r}
+i\lambda \sum_{\bf r}\Phi_{\bf r}} }\nonumber\\
&\propto & \int d\lambda e^{-2\pi^2\frac{\fluxvar}{\Phi_0^2} 
\sum_{\bf r}(w_{\bf r}+\lambda)^2} \quad\propto\quad e^{-S_2}\nonumber\\
&S_2& = 2\pi^2\frac{\fluxvar}{\Phi_0^2}
\left[\sum_{\bf r}w^2_{\bf r}-
\frac{1}{L^2}\left(\sum_{\bf r} w_{\bf r}\right)^2\right] 
\label{wr}
\end{eqnarray}
Thus we see that the action cost due to the current interaction is
proportional to a geometrical property of the world lines, similar to
an unoriented area, which has been termed the 
``Amperean area''\cite{wheatscho}:
\begin{equation}
{\cal A}_a=\left[\sum_{\bf r}w^2_{\bf r}-
\frac{1}{L^2}\left(\sum_{\bf r} w_{\bf r}\right)^2\right] 
\end{equation}
This geometrical interpretation of $S_2$ is particularly useful in the
numerical evaluation of this quantity.

If we are working with periodic boundary conditions, the geometrical
definition of $w_{\bf r}$ given above will not work because there is
an ambiguity in identifying which plaquettes are inside or outside a
loop on a torus. Nevertheless, we can still use the above analysis for
paths which do not wrap around the boundaries. (We will discuss
wrapping paths in the next section.) The only modification is that we need
a definition of the winding numbers which preserves Stokes' theorem:
$\oint {\bf a}({\bf x})\cdot d{\bf x}=\sum_{\bf r}w_{\bf r}\Phi_{\bf
  r}$. In the case of zero total flux, a suitable definition is:
$w_{\bf r}=\tilde\Phi^{-1}\oint [{\bf a}^0_{\bf r}({\bf x})-{\bf
  a}^0_{\bf R}({\bf x})]\cdot d{\bf x}$, where ${\bf a}^0_{\bf r}({\bf
  x})$ is the vector potential at ${\bf x}$ due to a test flux
$\tilde\Phi$ placed at plaquette ${\bf r}$, and ${\bf R}$ is an
arbitrary reference plaquette. Geometrically, this picks ${\bf R}$ to
be on the ``outside'' of any loop on the torus. The Amperean area as
defined above is independent of the choice of this plaquette, because
different choices amount to global changes in the winding numbers
({\em e.g.}, $w_{\bf r} \rightarrow w_{\bf r} + 1$) and the above
definition is invariant under such changes.

The effect of the gauge field on the particles is now clear. The
action $S_2$ suppresses world-line loops with large winding
numbers. Indeed, since $S_2$ is non-negative, it excludes all
configurations with finite Amperean area in the limit of infinite
$\fluxvar$. This suppression can be related to the original problem of
holes moving in a spin liquid with a slowly varying spin quantization
axis.  A hole moving in a loop comes back with a random phase due to
the locally fluctuating spin chiralities of the spin
background\cite{PN}. The random phase can be interpreted as arising
from a fictitious random flux.  World-line loops that enclose large
areas are strongly suppressed when averaged over random flux
distribution due to the destructive interference of the random
phases. Therefore, we expect that, in the presence of strong random
flux, the dominant contribution to the partition function comes from a
special kind of paths that do not ``see'' the random flux, {\em i.e.},
paths where $\int {\bf a}\cdot d{\bf x} =0$. These are ``retracing
paths'' where each traversal of a link on the lattice is retraced in
the opposite direction at some point in time
\cite{brinkman,oppermann}, and such paths have zero Amperean area.

A similar picture of retracing paths has been studied by Brinkman and
Rice\cite{brinkman} who studied a single hole in a Mott insulator
where the spins are treated classically. Indeed, studies of a single
particle in a strong random flux have yielded a density of states
nearly identical to that of the Brinkman-Rice problem\cite{zee,ziman,gavazzi}.
The Brinkman-Rice model gives a linear-$T$ resistivity at high
temperatures ($T>t$) but a constant scattering rate of order
$t$. Although we might expect this to be applicable to our model far
above the degeneracy temperature of the bosons, this behavior does not
extend down to the degenerate regime relevant to the present problem.

At boson densities of interest here and at low temperatures, Bose
statistics and particle exchange are important; they can give rise to
behavior different from the single-particle Brinkman-Rice result. We
shall look at the effect of the gauge field on the quantum exchanges
among bosons more carefully in section IV-c. For now, we point out
that, even in the presence of strong gauge-field fluctuations, the
bosonic nature of the particles cannot be ignored because the
particles can form long exchange cycles that retraces themselves so
that an individual boson does not have to retrace its own path.  This
is an important consideration at low temperatures where the
imaginary-time paths are long allowing for a strong degree of particle
exchange. Although the system can be highly degenerate at low
temperatures, we shall now argue that these ``Brinkman-Rice bosons''
remain normal at all finite temperatures, due to interactions with the
fluctuating gauge fields.

\subsection{Destruction of superfluidity}

We will now discuss the effect of the gauge field fluctuations on the
superfluidity of the Bose system. We will see, as in the previous
section, that this can be understood in terms of the geometrical
properties of the boson world lines.

A neutral Bose system with short-range interaction in two dimensions
is a superfluid below Kosterlitz-Thouless temperature $T_{\rm
KT}$. The onset of superfluidity at $T_{\rm KT}$ is caused by the
binding of vortex-antivortex pairs in the Bose fluid so that vortex
motion does not cause phase slips across the system. An essential
ingredient of the existence of the superfluid phase is a long-range
logarithmic attraction between the vortices and the antivortices. A
single vortex costs infinite energy in an infinite system $E_v =
(\pi\rho_s/m)\log (L/a)$ where $a$ is a short distance cutoff ($\sim$
vortex core radius) and $\rho_s$ is the superfluid density. Therefore
single vortices cannot exist at low temperatures. Nevertheless, the
proliferation of free vortices is possible above $T_{\rm KT}$ because
this provides a gain in entropy which also scales as $\log
L$. However, in a charged Bose system, screening currents causes the
vortex interaction to be short-ranged. In our problem, the vortex
interaction becomes exponentially weak at distances beyond
$\lambda_P=[T/2\rho_st\fluxvar]^{\frac{1}{2}}\Phi_0$, which can be
interpreted as a penetration depth of the Bose fluid. Now, the
creation of a single vortex costs a finite amount of
energy\cite{nagGinz,feigel} $E_v= (\pi\rho_s/m)\log (\lambda_P/a)$. This
no longer compensates the entropic gain from vortex-antivortex
unbinding, and so we do not expect to see a sharp phase transition of
the Kosterlitz-Thouless type at finite temperatures.

One might still expect that there is a crossover temperature scale
below which the vortex density will be sufficiently low that the Bose
system would have strong diamagnetic response. A rough estimate of
this temperature scale using a Boltzmann weight for the vortex density
gives a large value for this crossover
temperature\cite{feigel}. However, we will see later that, in the
presence of strong gauge fluctuations, the diamagnetic response of the
bosons remains small.

\begin{figure}[hbt]
\epsfxsize=\columnwidth\epsfbox{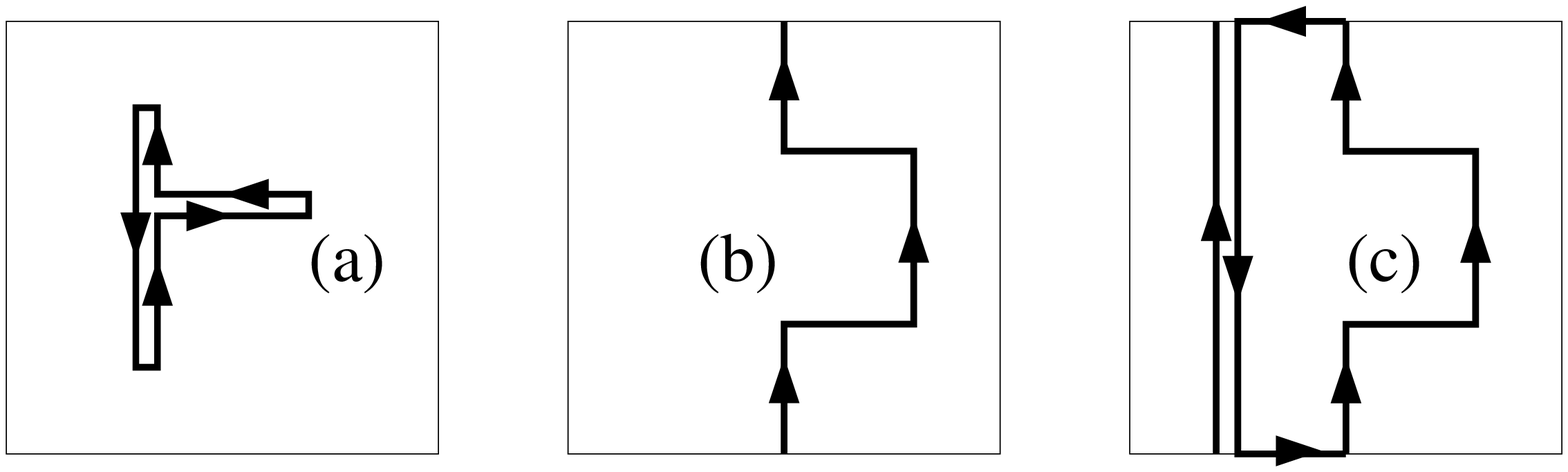}
\caption{(a) Projection of a world line onto the $xy$ plane shows a
retracing path. (b) A wrapping path. (c) Decomposition of (b) into a
reference path and a non-wrapping path.}
\label{figwrap}
\end{figure}

To understand the suppression of superfluidity specifically in our
model, we turn to the path-integral formulation of the problem with
periodic boundary conditions ({\em i.e.}, on a torus). Ceperley and
Pollock\cite{ceppol} have shown that superfluidity is associated with
the existence of long world-line cycles which wrap around the torus.
The superfluid density is given by
\begin{equation}
n_s=\frac{\langle {\bf W}^2 \rangle}{4\beta t}
\label{sfdef}
\end{equation}
where $W_x$ $(W_y)$ is the number of times the boson world lines wrap
around the torus in the $x$ $(y)$ direction. In other words, ${\bf
W}=\sum_{\alpha=1}^{N}\int_0^{\beta}d\tau\,\dot{\bf x}_{\alpha}/L$.
In the presence of gauge fields, superfluidity is destroyed by the
same mechanism that causes the Brinkman-Rice behavior: wrapping
configurations pick up random phases, and should be suppressed by
destructive interference on averaging over the gauge field. The number
of plaquettes whose random fluxes contribute to the phase picked up by
a wrapping path should increase with increasing system size. For a
large enough system, one might expect this phase to be totally
random. We therefore expect this suppression to be very strong. For
instance, one can evaluate $S_2$ for a straight-line path which wraps
around the torus in the $y$-direction. To do so, we use Eq.\
(\ref{currint}) instead of Eq.\ (\ref{wr}) because the geometrical
interpretation of $S_2$ in terms of winding numbers is not applicable
for wrapping paths. We find that such a path gives
\begin{equation}
S_2 = \frac{W_y^2}{2\beta} \sum_{k_x\neq 0,k_y=0}\!\!\! D({\bf k},0) \simeq
2\pi^2 W_y^2 L^2 \frac{\fluxvar}{\Phi_0^2}.
\end{equation}
To compute $S_2$ for a more general path with wrapping $W_y$, one can
break it down into a wrapping path with the same wrapping number and a
non-wrapping path (Fig.~\ref{figwrap}). ($S_2$ will consist of the
contributions of the wrapping paths and non-wrapping paths separately,
as well as a cross-term between the two paths.) We argue that $S_2$
diverges for all wrapping paths in the thermodynamic limit, and so
superfluidity is destroyed at all finite temperatures.

It should be noted that we are studying a gauge model where the
uniform part of {\bf a} is set to zero. We may alternatively work with
a model without this gauge fixing. With periodic boundary conditions,
this model allows an arbitrary Aharonov-Bohm (AB) flux through the
torus. This flux is related to the phase of the product of RVB
parameters ($\xi_{ij}$) along a (Wilson) loop which wraps around the
torus. If we average over this AB flux assuming a uniform
distribution, we would find that all wrapping paths are strictly
prohibited and $n_s=0$ at {\em all\/} temperatures even for samples of
finite size. We will not impose such a drastic condition on the
wrapping paths in this work.

We can also ask whether long-range order exists in the Green's
function for the bosons. The Green's function itself $\langle
b^{\dagger}({\bf r})b(0)\rangle$ is not gauge invariant, and would vanish
on averaging over different gauges.  However, we can
study the Green's function in a fixed gauge, for example the
transverse gauge $\nabla\cdot{\bf a}=0$. In fact, one can write a
gauge-invariant analogue correlation function which coincides with the
Green's function in the transverse gauge\cite{fradbook}:
\begin{eqnarray}
G({\bf r}) &=& \langle b^{\dagger}({\bf r})b(0)
  \rangle_{\nabla\cdot{\bf a}=0}\nonumber\\
&=& \langle b^{\dagger}({\bf r})b(0)e^{-i\int d^2{\bf r'}f({\bf r}')
  \nabla\cdot{\bf a}({\bf r}')}\rangle 
\label{greendef}
\end{eqnarray}
where $\nabla_{{\bf r}'}^2 f({\bf r}') = \delta({\bf r}'- {\bf
r})-\delta({\bf r}')$. In the path-integral representation, the
evaluation of $G$ involves a world line originating at site ${\bf r}$
and a world line terminating at site $0$ at the same point in
imaginary time. Note that this quantity coincides with the Green's
function in the Coulomb gauge. Consider now the phase factor
$\sum_\alpha \int {\bf a}\cdot d{\bf x}_\alpha$ picked up by the world
lines $\{{\bf x}_{\alpha}\}$ in the evaluation of the Green's function
$G({\bf r})$ in this gauge. The random flux $\Phi_{\bf R}$ at a distant
plaquette ${\bf R}$ (with $R\gg r$) has a contribution of magnitude
$\Phi_{\bf R}/R$ to the vector potential at a point $Q$ near $0$ and
${\bf r}$. The sum of the contributions to the vector potential at $Q$
due to the random fluxes at radius $R$ from the origin is a random
vector with a mean squared magnitude of $2\pi R \times
(\fluxvar/R^2)\sim\fluxvar/R$. This analysis is valid for all fluxes
which are at a distance $R>r$.  Integrating over the contribution of
such fluxes, the variance of the magnitude of the vector potential at
$Q$ scales as $\fluxvar\log(L/r)$.  Summing over all $Q$ near 0 and
${\bf r}$, we obtain a random phase with a divergent variance:
$\fluxvar r^2\log(L/r)$. Thus, averaging over the distant fluxes for
these sites, one obtains a suppression factor of $\exp[-\fluxvar r^2
g(r/L)]$ where $g(x)\sim \log 1/x$ for small $x$.  This can be
interpreted as a binding potential for the end points of $G({\bf r})$.
We therefore do not find long-range order in this quantity because of
the destructive interference of the random phases due to distant
fluxes.


We will now present numerical evidence for the suppression of
superfluidity below the degeneracy temperature $T_{\rm D0}$ of the
system. A measure of the degeneracy temperature is the
Kosterlitz-Thouless temperature of the system at zero flux. We make
use of the observation of Ceperley and Pollock\cite{ceppol2d} that the
the probability of bosons to participate in the multi-particle
exchange is about $\frac{1}{2}$ at Kosterlitz-Thouless transition. In
other words, the probability $P_1$ that a boson is in an exchange
cycle of length 1 is about $\frac{1}{2}$. We estimate that, for our
lattice bosons with density $n_b=0.25$ and on-site interaction $U=4t$,
the degeneracy temperature $T_{\rm D0}=1.1t$. (For strong on-site
repulsion, $T_{\rm D0}$ is not particularly sensitive to the value of
$U$, {\em e.g.}, $T_{\rm D0}=0.9t$ for $U=16t$.)

\begin{figure}[bt]
\epsfxsize=0.9\columnwidth
\epsfbox{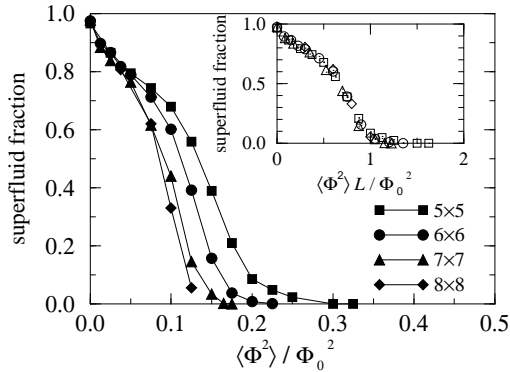}
\caption{Superfluid density {\em vs.}\ $\fluxvar$ for different system sizes
  at $\beta t=6$. Inset: a scaling plot suggests that superfluidity
  vanishes at $\langle\Phi^2\rangle_{\rm c}\sim 1/L$.}
\label{figsf}
\end{figure}

We have measured, using Eq.\ (\ref{sfdef}), the superfluid fraction
$n_s/n_b$ at $T=t/6$ with $U=4t$ for a range of flux variances and for
systems up to 8$\times$8 in size (Fig.~\ref{figsf}). We see that the
superfluid fraction decreases with increasing system size. In fact,
the superfluid fraction as a function of $\fluxvar L$ collapses onto a
single curve (Fig.\ (\ref{figsf} inset), indicating that
$n_s(L,\beta,\fluxvar)= f(L\fluxvar,\beta)$. Since $f(x,\beta)\to 0$
as $x\to\infty$, we see that an arbitrarily small random magnetic flux
would destroy superfluidity in the thermodynamic limit.  In the
language of the renormalization group, this shows that the scattering
by gauge fields is a relevant perturbation at finite temperature.

\subsection{World line geometry in the normal phase}

Having established that our system remains normal at low temperatures,
we will now examine the geometry of the world lines in this normal
phase in the presence of strong gauge fluctuations. In particular, we
will look at the effect of the gauge fields on quantum exchange and
imaginary-time diffusion. These are mutually related: imaginary-time
diffusion over large distances aids quantum exchange among particles
and quantum exchange facilitates imaginary-time diffusion.  For
example, in a dissipative model of bosons coupled to external heat
bath, a slow logarithmic imaginary-time diffusion is expected to
suppress quantum exchange very strongly, resulting in an incoherent
liquid even at zero temperature\cite{wheathong,wheatstat}. In our
case, the bosons are elastically scattered by the gauge fields. We
find that the gauge fields have less dramatic effects on quantum
exchange and imaginary-time diffusion.

We have shown that the world lines retrace themselves in the presence
of random flux. One might expect that, compared with the case of zero
flux, this would reduce the distance traveled by the particles in the
imaginary-time interval $\beta$ before their paths must return to some
permutation of their starting positions. This should slow down the
imaginary-time motion of the bosons as well as reduce the
probabilities for exchange. We find that this is indeed the case.

We first look at the exchange probabilities $P_i$ of a particle
participating in an exchange cycle of $i$ bosons. As before, we may
deduce a degeneracy temperature $T_{\rm D}$ from the probability
$(1-P_1)$ for a particle to be involved in particle 
exchange\cite{wheatcoh}. This
degeneracy temperature is reduced compared to the case of zero flux.
For $U=4t$ at quarter filling, we find that the zero-flux degeneracy
temperature $T_{{\rm D}0}=1.1t$ is reduced to $T_{\rm D}=0.5t$ at
$\fluxvar=0.5\Phi_0^2$. 
At $\frac{1}{6}$-filling, it is reduced from $T_{{\rm D}0}=0.8t$ to
$T_{\rm D}=0.34t$.
A finite $T_{\rm D}$ does not imply Bose condensation at a finite
temperature. Indeed, one cannot deduce a superfluid transition by
examining the exchange probabilities. Remarkably, in the degenerate
regime below $T_{\rm D}$, the exchange probabilities for the cases of
$\fluxvar=0$ and $0.5\Phi^2_0$ are nearly identical (see
Table~\ref{tab1}). In this temperature regime, a particle is equally
likely to participate in an exchange cycle of any size: $P_1\simeq P_2
\simeq \ldots \simeq P_N \simeq 1/N$.

\begin{figure}[htb]
\epsfxsize=0.9\columnwidth\epsfbox{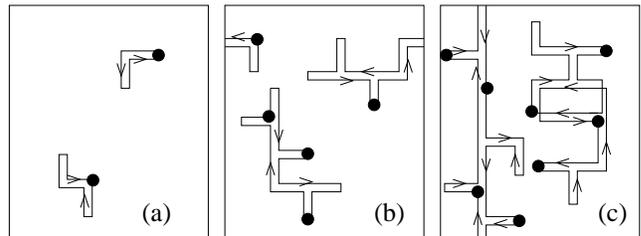}
\caption{Schematic world-line cycles which retrace when projected onto
the $xy$-plane.  Solid circles denote boson positions at $\tau=0$. (a)
Each boson retraces its own path; (b) Exchange cycles with more than
one boson retrace their own paths; (c) Two exchange cycles can retrace
each others paths, and two wrapping paths can retrace each other to give
zero total wrapping around the boundaries.}
\label{cycleretrace}
\end{figure}

We can gain a qualitative understanding of the low-temperature
exchange probabilities, by examining how the suppression of Amperean
area by $S_2$ affects the geometry of the world-line
configurations. When there is significant quantum exchange, individual
bosons do not have to retrace their own paths in order to minimize the
total Amperean area of the world-line configuration of all the
bosons. Instead, one might minimize the Amperean area of each
world-line loop formed by several bosons in the same exchange
cycle. We find that this is not the entire situation at sufficiently
low temperatures. Below $T_{\rm D}$, the different world-line loops
have strong overlap. We find that different cycles retrace each
others' paths. (See Fig.~\ref{cycleretrace}.) Thus, although the gauge
fields have a drastic effect on the {\em total\/} area enclosed by all
the boson world lines, individual world-line cycles may enclose large
areas. One might therefore expect that some aspects of the world-line
geometry, which are insensitive to the total area, may indeed be
very similar to the case of zero flux.

\begin{figure}[htb]
\epsfxsize=0.9\columnwidth\epsfbox{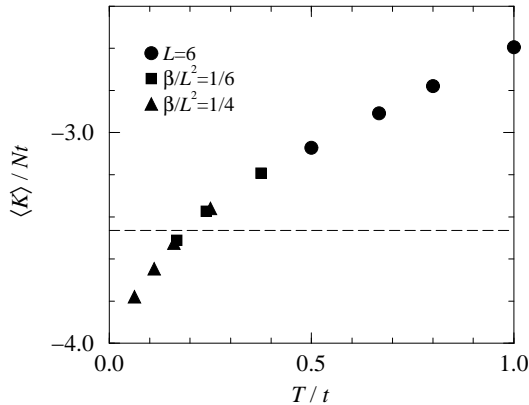}
\caption{Kinetic energy per particle as a function of temperature. 
  Dashed line marks the Brinkman-Rice band edge for the
  single-particle problem. $\fluxvar=0.5\Phi_0^2$ and $U=4t$.}
\label{figkin}
\end{figure}

The observation that individual particles do not have to retrace their
own paths suggests that they could diffuse a greater distance than in
the single-particle case. One should see a reduction in the kinetic
energy $\langle K\rangle$ of the particles compared to the
Brinkman-Rice theory\cite{brinkman}. This is indeed the case. (See
Appendix for a discussion of the measurement of the kinetic energy.) A
single particle with retracing paths has a band edge at $-2\sqrt{3}t$
rather than $-4t$. In our system, the kinetic energy per particle goes
below the Brinkman-Rice band edge at low temperatures, approaching
$-4t$ roughly linearly in temperature (Fig.~\ref{figkin}). Thus, we
see that the strong gauge fluctuations do not have a large effect
on some aspects of the world-line geometry ({\em e.g.}, exchange
probabilities) while having a dramatic influence on others ({\em e.g.},
superfluidity).

\begin{figure}[htb]
\epsfxsize=0.9\columnwidth\epsfbox{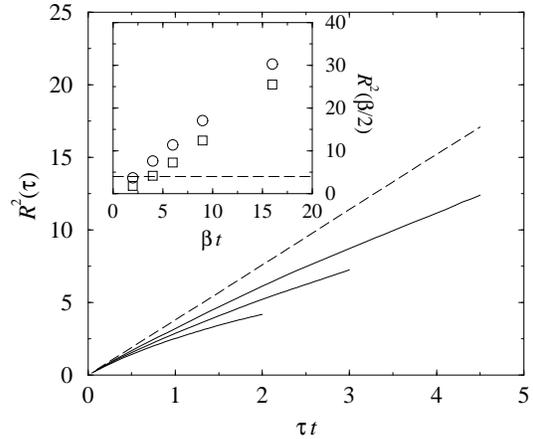}
\caption{Single-particle diffusion $R^2(\tau)=\langle [{\bf
  x}(\tau)-{\bf x}(0)]^2\rangle$ in imaginary time for $0 < \tau <
  \beta/2$. Solid lines: strong random flux with
  $\fluxvar=0.5\Phi_0^2$ at $\beta t=4,6,9$. Dashed line: zero flux at
  $\beta t=9$. Inset: $R^2(\beta/2)$ for zero flux ($\circ$) and
  $\fluxvar=0.5\Phi^2_0$ ($\Box$); dashed line marks the squared
  interparticle spacing.}
\label{diffusion}
\end{figure}

Let us now examine the imaginary-time motion of the particles in more
detail. Ideal bosons are diffusive in imaginary time at all
temperatures, {\em i.e.}, the mean-squared displacement of particle
$\alpha$ is linear in imaginary time $\tau$: $R^2(\tau)=\langle [{\bf
  x}_\alpha(\tau)-{\bf x}_\alpha(0)]^2\rangle = 4t \tau$ for
$0<\tau<\beta/2$. With repulsive interactions, there is an increase in
the effective mass of the particle, {\rm e.g.}, for $U=4t$ at quarter
filling, we find $t \rightarrow t^* = 0.95t$.
In the presence of random magnetic flux, the imaginary-time diffusion
is slowed down, and the mean-squared displacement $R^2(\tau)$ is no
longer linear in $\tau$ at all temperatures. Fig.~\ref{diffusion}
shows our results for the (superfluid) zero-flux case at temperature
$\beta t=9$ and the case of strong random flux at $\beta t=4,6,9$.
Since we are working with periodic boundary conditions, we have used
the definition: $R^2(\tau)=\langle [\int_0^{\beta/2}\!\!\dot{\bf
  x}_\alpha(\tau)d\tau]^2\rangle$. We can see that, whereas
$R^2(\tau)$ has significant downward curvature at $\beta t=2$, it
becomes closer to diffusive behavior as the temperature is lowered.
However, we are unable to reach the asymptotic regime where the
particle has traveled far on the scale of the interparticle spacing
over a time period of $\beta/2$ (see Fig.~\ref{diffusion} inset).

\begin{figure}[thb]
\epsfxsize=0.9\columnwidth\epsfbox{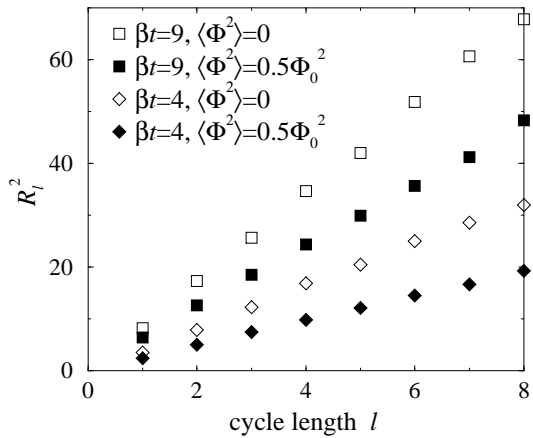}
\caption{Cycle sizes $R_l^2$ as a function of cycle length $l$ for a 
6$\times$6 lattice with 9 particles.}
\label{figcycle}
\end{figure}

In order to study the long-time behavior, we can examine the size of
the world-line exchange cycles. A cycle where the world lines of $l$
particles $\{{\bf x}_1,\ldots,{\bf x}_l\}$ form a loop can be roughly
regarded as a particle traveling over a time interval of $l\beta$.
Thus, the possibility of exchange means that a world-line cycle can
travel large distances compared with an individual boson. In a
system with periodic boundaries, the size $R_l$ of the cycle is
defined by:
\begin{eqnarray}
R_l^2 
&=& \left\langle \left[\int_0^{\beta/2}\!\!\dot{\bf x}_{\frac{l+1}{2}}\,d\tau
  +\sum_{\alpha=1}^{(l-1)/2}
  \int_0^{\beta}\!\!\dot{\bf x}_\alpha\,d\tau\right]^2\right\rangle
        \quad l{\rm\ odd}, \nonumber\\
&=& \left\langle\left[\sum_{\alpha=1}^{l/2}
  \int_0^{\beta}\!\!\dot{\bf x}_\alpha\,d\tau \right]^2\right\rangle
        \quad l{\rm\ even}.
\end{eqnarray}
For ideal bosons, $R_l^2$ should equal $R^2(\tau=l\beta/2)$ at inverse
temperature $l\beta$, and therefore should scale linearly with $l$.
Fig.~\ref{figcycle} shows $R_l^2$ for a 4$\times$4 lattice with 9
particles. We have measured only cycles which do not have a net
wrapping number around the periodic boundaries so that we do not have
contributions from cycles with different topologies. We see that
$R^2_l$ is linear in $l$ for the cases of zero flux and strong random
flux, although the slope of the case with the strong random flux is
reduced substantially. This demonstrates that the imaginary-time
motion of the bosons is diffusive at long distances.

These results indicate that we are probing an unconventional phase of
a Bose liquid. Although the system remains normal, many aspects of the
imaginary-time motion of the particles in the degenerate regime
resemble that of a neutral Bose liquid which is a superfluid in such
temperatures. In subsequent sections, we shall study the physical
properties of this ``strange metal'' and discuss the relevance to
the normal state of the cuprate superconductors.

\section{Transport and optical conductivity}

\begin{figure}[bth]
\epsfxsize=0.9\columnwidth\epsfbox{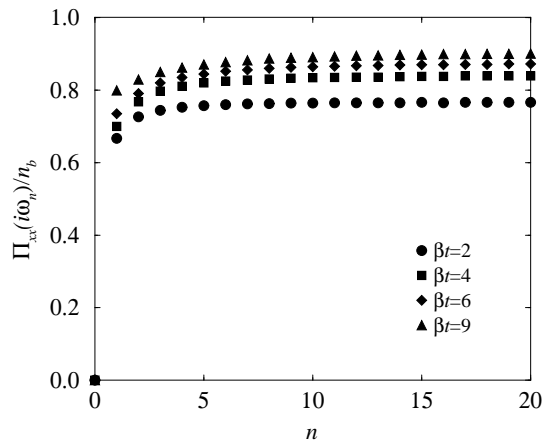}
\caption{Current correlation function $\Pi_{xx}(i\omega_n)$ for 
  a 6$\times$6 lattice with 9 bosons with $\fluxvar=0.5\Phi_0^2$ and
  $U=4t$.}
\label{sfw}
\end{figure}

In this section, we will present our quantum Monte Carlo (QMC) results on
longitudinal transport for this strange Bose metal. To obtain the
conductivity of the system, we measure its imaginary-time analogue
$\sigma_{\alpha\beta}(i\omega_n)$ in our quantum Monte Carlo
simulation:
\begin{eqnarray}
\sigma_{\alpha\beta}(i\omega_n)&=&\frac{1}{|\omega_n|}
{\Pi}_{\alpha\beta}(i\omega_n)  \\
{\Pi}_{\alpha\beta}(i\omega_n)&=&\int_0^{\beta}\langle
j_{{\bf q}=0}^{\alpha}(\tau)j_{{\bf q}=0}^{\beta}(0)\rangle
e^{i\omega_n\tau}d\tau,
\label{kubocond}
\end{eqnarray}
where ${\bf j}_{\bf q}(\tau)=\sum_{\bf r}{\bf j}_{\bf r}(\tau) e^{
i{\bf q}\cdot{\bf r}}$ and ${\bf j}_{\bf r}(\tau)=\sum_{\alpha}
\delta({\bf r}-{\bf x}_{\alpha}(\tau))\frac{d{\bf x}_{\alpha}}{d\tau}$ 
is the gauge-invariant current (Fig.~\ref{sfw}).

The imaginary-time measurements are related to the
real-time conductivity $\sigma(\omega)\equiv\sigma_{xx}(\omega)$ by:
\begin{equation}
-\frac{1}{2L^2}
\langle {\bf j}_{{\bf q}=0}(\tau)\cdot{\bf j}_{{\bf q}=0}(0)\rangle =
        \int_{-\infty}^{\infty}\!\! 
        \frac{\omega e^{-\omega\tau}\sigma(\omega)}{1-e^{-\beta\omega}}
        \frac{d\omega}{\pi}.
\label{kubo}
\end{equation}
Deducing dynamical properties (such as conductivity) from
imaginary-time data is in general an ill-posed problem. Several
approximate methods are often used in the context of QMC studies. A
simple method, which has been used in the study of the
superfluid-insulator transition\cite{girvin,batrouni}, is to fit
$\sigma(i\omega_n)$ to a simple functional form, such as the Drude
form $\sigma(i\omega_n)= \sigma_0/(1+|\omega_n|\tau_{\rm tr})$.  More
generally, one can use a Pad\'e approximant to fit an arbitrary number
of poles and zeroes:
\begin{equation}
\sigma(z) = \frac{a_0+a_1 z + 
\dots a_{N_n} z^{N_n}}{b_0+b_1 z + \dots b_{N_d} z^{N_d}}.
\label{pade}
\end{equation}
This approach is particularly suitable if the scattering rate
$1/\tau_{\rm tr}$ (or the position of the pole closest to the origin
in (\ref{pade})) is large compared to the temperature at low
temperatures. This is however not the case in our problem. In our
system, $\Pi_{xx}(i\omega_n)$ is nearly constant as a function of
$n$ for finite $n$ even at low temperatures, suggesting that
$1/\tau_{\rm tr}$ is proportional to $T$. (Note that
${\Pi}_{xx}(n=0)=0$ in the limit of strong random flux because paths
which wrap around the torus are strongly suppressed.)

We have calculated the conductivity by numerical analytic continuation
using the maximum-entropy (MaxEnt) method\cite{skilling,gubernatis}.
Eq.\ (\ref{kubo}) takes the form of a linear integral equation:
\begin{equation}
d(\tau)=\int\! K(\tau,\omega) r(\omega) \,d\omega,
\end{equation}
where $K(\tau,\omega)$ is the kernel relating the imaginary-time data
$d(\tau)$ to the corresponding response function $r(\omega)$. In our
QMC simulations, $d(\tau)$ is measured at discrete points
$\tau_l=l\Delta\tau$ with mean $\overline{d}_l$. The errors for the
time points $l$ and $m$ are correlated with a covariance matrix
$C_{lm}=\langle (d_l-\overline{d}_l)(d_m-\overline{d}_m) \rangle$.
The MaxEnt method finds an estimate of $r(\omega)$ as the function
${\hat r}(\omega)$ which maximizes the functional:
\begin{equation}
 \phi[{\hat r}(\omega);\alpha] = -\frac{1}{2} \chi^2 + \alpha S, 
\label{maxent}
\end{equation}
where $\chi^2$ is the goodness of fit
\begin{equation}
\chi^2= \sum_{l,m}  (D_l-\overline{d}_l) [C^{-1}]_{lm} (D_m-\overline{d}_m)
\end{equation}
with $D_l=\int d\omega K(\tau_l,\omega){\hat r}(\omega)$, and the
``entropy'' $S$ is measured with respect to a given default model (or
measure) $m(\omega)$:
\begin{equation}
S=\int d\omega \left[
{\hat r}(\omega)-m(\omega)-{\hat r}(\omega)
\log\frac{{\hat r}(\omega)}{m(\omega)}\right].
\end{equation}
The variable $\alpha$ in Eq.\ (\ref{maxent}) is a regularization
parameter controlling the competition between the smoothness and the
goodness of the fit, and $\phi$ is also maximized with respect to
it\cite{wolfgang}. We have chosen the default model $m(\omega)$ to be
a constant in order not to build in any bias. Our results are not
sensitive to this choice. Details of the MaxEnt method are given in
Refs.\ \onlinecite{skilling,gubernatis,wolfgang}.

One can check the results of the MaxEnt inversion using relevant sum
rules. In the case of conductivity, we have used the sum rule
\begin{equation}
  \int_{0}^{\infty} \sigma(\omega) d\omega =
-\frac{\pi}{4}\frac{\langle K\rangle}{L^2}.
\label{csum}
\end{equation}
which is the lattice version of the more familiar form in the
continuum: $\int_0^{\infty}\sigma(\omega)d\omega=\pi n_b/2m$. In our
MaxEnt results, this sum rule is obeyed to within $3\%$ error.  In
order to obtain reliable data for the MaxEnt inversion, we have worked
with a fine discretization in imaginary time ($t\Delta\tau\le 0.1$).
For the lowest temperatures ($T < 0.4t$), we worked at fixed
$\Delta\tau$ and $\beta / L^2$. We chose $\beta \propto L^2$ to
control the finite-size effects because of the imaginary-time motion
of the bosons is roughly diffusive, as discussed above. Our results
are in fact not very sensitive to this choice, indicating that
finite-size effects are small. For instance, the values of resistivity
at $\beta t=4$ and $n_b=1/4$ for $4\times 4$ and $6\times 6$ are
similar within statistical error.

\begin{figure}[thb]
\epsfxsize=0.9\columnwidth\epsfbox{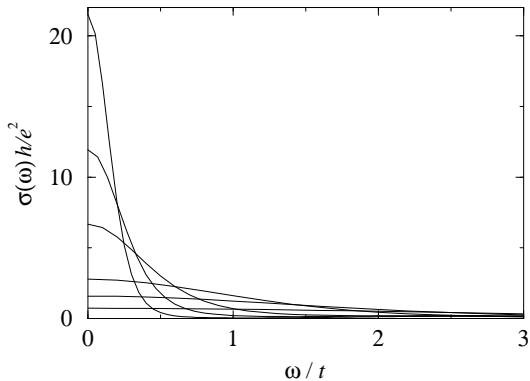}
\caption{Optical conductivity for 6$\times$6 lattice with 9 bosons at
$\beta t =$9,6,4,2,1,0.5. $\fluxvar=0.5\Phi^2_0$ and $U=4t$.}
\label{figac}
\end{figure}

We find that $\sigma(\omega)$ consists of a single Drude-like peak
(Fig.~\ref{figac}). Since this peak exhausts the sum rule
(\ref{csum}), its spectral weight is proportional to $-\langle
K\rangle$. This spectral weight has a weak temperature dependence in
this temperature range because, as already discussed, the kinetic
energy approaches $-4t$ per particle as the temperature is
lowered. This should be contrasted with the Brinkman-Rice result
\cite{brinkman} for non-degenerate particles ($T \gg t$) where the
weight under $\sigma(\omega)$ decreases as $\langle -K\rangle\sim
T^{-1}$.

\begin{figure}[hbt]
\epsfxsize=0.9\columnwidth\epsfbox{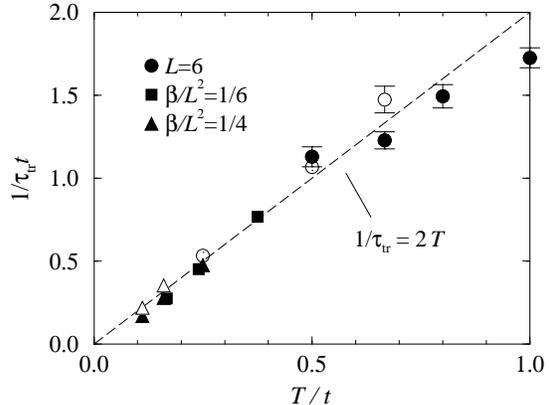}
\caption{Scattering rate $1/\tau_{\rm tr}$ as a function of temperature.
Solid(hollow) symbols correspond to a boson density of $n_b$=1/4(1/6).
$\fluxvar=0.5\Phi^2_0$ and $U=4t$.}
\label{figtr}
\end{figure}

\begin{figure}[hbt]
\epsfxsize=0.9\columnwidth\epsfbox{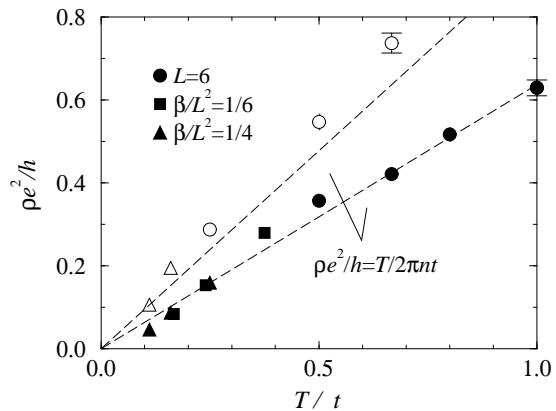}
\caption{Resistivity as a function of temperature. 
Solid(hollow) symbols correspond to a boson density of $n_b$=1/4(1/6).
$\fluxvar=0.5\Phi^2_0$ and $U=4t$.}
\label{figresist}
\end{figure}

The width of $\sigma(\omega)$ gives a transport scattering rate
consistent with: $1/\tau_{\rm tr} = \zeta k_{\rm B}T$ with $\zeta=1.8
- 2.2$ (Fig.~\ref{figtr}). This result has been obtained for two
densities $n_b=1/4$ and $1/6$ so that this scattering rate appears to
be independent of density. Again this differs from the Brinkman-Rice
result where $1/\tau_{\rm tr}$ is a constant of order $t$ (as one
begins to see at the highest $T$ in Fig.~\ref{figtr}). The
resistivity $\rho$, given by the peak height, is consistent with a
linear temperature dependence of $\rho e^2/h = (1/2\pi n_b) T/t$ for
$T < 2t$ (Fig.~\ref{figresist}). We estimate a statistical error of
5\% for $\rho$ by examining fluctuations due to statistical errors in
the measurement of the current correlation function. There are also
systematic errors due to the smoothing of structures.

There appears to be a systematic deviation from the linear-$T$
behavior below $T=0.3t$, in particular in the case of quarter filling.
This deviation is stronger for $\rho$ than for $1/\tau_{\rm tr}$. The
difference can be attributed to the $T$-dependence of the Drude weight
discussed above which should affect the resistivity but not the
relaxation time. We speculate that the deviation from linearity at the
lowest temperatures may indicate the approach to zero-temperature
critical behavior. This is beyond the scope of this paper.

Our resistivity agrees, to within a factor of 2, with Jakli\v{c} and
Prelov\v{s}ek \cite{jaklic} who provided an approximate
diagonalization of the $t$-$J$\ model on 4$\times$4 lattices and found
a Drude peak with width $2T$. They also found a broad background, and
interpreted it with a frequency-dependent scattering rate
$\tau(\omega)$. Indeed, some authors have interpreted the experimental
optical conductivity as possessing a power-law tail and emphasized its
importance\cite{anderson2}.
This incoherent part of the conductivity is absent from our
boson model, and may be due to inelastic scattering of the bosons with
the gauge field or, more generally, with the fermionic degrees of
freedom. 

\section{Magnetic response}

We now discuss the response of this degenerate Bose liquid to a weak
external magnetic field perpendicular to the plane. In the absence of
the random magnetic fields, a Bose liquid has a strong diamagnetic
response as the temperature is lowered towards the transition to a
superfluid when it develops a Meissner response. We argue here that
the linear response of the system to a magnetic field is strongly
suppressed by the gauge fluctuations. Qualitatively, this can be again
understood by examining the world-line configurations. We have already
demonstrated that the partition function is dominated by world-line
paths which are unaffected by the internal gauge fields $\sum_\alpha
\int\!{\bf a}\cdot d{\bf x}_\alpha=0$ for any ${\bf a}$. These
configurations are therefore also unaffected by any external magnetic
fields. Thus, we see that the system has a vanishing linear
response to magnetic fields in the limit of strong gauge
fluctuations. For the sake of completeness, we will now discuss more
quantitatively the magnetic response of the system. Relevant physical
quantities are the diamagnetic susceptibility $\chi_{\rm B}$, the Hall
coefficient $R_{\rm H}$, and the magnetoresistance $\Delta\rho /\rho$.

Consider first the diamagnetic susceptibility. On the infinite plane,
in the presence of a weak external field $H$, each world-line
configuration picks up an extra factor of $\exp [-i\sum_\alpha {\bf
A_{\rm ext}}\cdot d{\bf x}_\alpha] = \exp [-iH {\cal A}_0]$ where
curl${\bf A_{\rm ext}}=H$, and ${\cal A}_o=\sum_{\bf r} w_{\bf r}$ is
the oriented area of the configuration. (In this section, we will use
units where $\Phi_0=2\pi$.) Expanding this in a Taylor expansion, one
can write the partition function $Z(H)$ as:
\begin{eqnarray}
  Z(H) &=& \int\!{\cal D}\{{\bf x}\} 
  \left(1\!-\!iH{\cal A}_o\!-\!\frac{1}{2}H^2{\cal A}_o^2\right)
  e^{-S_{\rm eff}} \nonumber\\
  &=& Z(0)\left(1 - \frac{1}{2}H^2\langle{\cal A}_o^2\rangle\right).
\end{eqnarray}
where ${\cal A}_o$ is the oriented area of a world-line configuration
and $\langle\ldots\rangle$ denotes an average for the system at $H=0$.
We have assumed here that the external magnetic field $H$ has
negligible effect on the spectrum of the gauge fluctuations. The
diamagnetic susceptibility is given by:
\begin{equation}
  \chi_{\rm B} = \frac{1}{\beta}\frac{\partial^2 \ln Z}{\partial H^2} = 
  \frac{4\pi^2T}{\Phi_0^2}\langle{\cal A}_o^2\rangle.
\end{equation}
Since ${\cal A}_a > {\cal A}_o$ by definition, we can see that, when
the gauge fluctuations are strong so that configurations with zero
Amperean area dominate, the system has no diamagnetic response, as
previously suggested in Section IV B.

It should be noted that, with periodic boundary conditions, the total
flux penetrating the torus is quantized in units of the flux quantum.
One should use replace $\langle{\cal A}_o^2\rangle$ by $4\langle\sin^2
[H_0{\cal A}_o/2]\rangle/H_0^2$ where $H_0=\Phi_0/L^2$ is the smallest
uniform field allowed in a torus of size $L$. Moreover, as in the case
for the Amperean area, a geometrical interpretation of the phase
factor $\int {\bf A_{\rm ext}}\cdot d{\bf x}$ is not possible for
paths which wrap around periodic boundaries. However, these wrapping
configurations are strongly suppressed in the case of strong random
flux and should give negligible contribution to the susceptibility.

Magnetotransport properties can be written in terms of the
conductivity tensor $\sigma_{\alpha\beta}^H$:
\begin{eqnarray}
  R_{\rm H}&\approx&\sigma_{xy}^H/(H\sigma_{xx}^2) \\
 \Delta\rho/\rho&\approx&-\Delta\sigma_{xx}/\sigma_{xx}
\end{eqnarray}
where $\sigma_{xx}=\sigma_{xx}^{H=0}$, and $\Delta\sigma_{xx}\equiv
\sigma_{xx}^H-\sigma_{xx}$
To obtain the conductivity tensor, we need the current-current
correlator $\langle j^{\alpha}(\tau)j^{\beta}(0)\rangle_H$. Expanding
again in a Taylor series in $H$, one obtains the correlator:
\begin{eqnarray}
 \langle j^{\alpha}j^{\beta}\rangle_H=\frac{\int\!{\cal D}
 \{{\bf x}\} j^{\alpha} j^{\beta}\left(1\!-\!iH{\cal A}_o\!-\!\frac{1}{2}
 H^2{\cal A}_o^2\right)
 e^{-S_{\rm eff}}}{\int\!{\cal D}\{{\bf x}\} 
\!\left(1-iH{\cal A}_o-\frac{1}{2}H^2{\cal A}_o^2\right)
 e^{-S_{\rm eff}}} \nonumber\\ 
 = \frac{\langle j^{\alpha} j^{\beta}\rangle - 
 iH\langle j^{\alpha} j^{\beta}{\cal A}_o\rangle
 -\frac{1}{2}H^2\langle j^\alpha j^\beta{\cal A}_o^2\rangle+\cdots}
{1-\frac{1}{2}H^2\langle{\cal A}_o^2\rangle + \cdots}
\end{eqnarray}
Terms such as $\langle
j^x j^x{\cal A}_o\rangle$, $\langle j^x j^y{\cal A}_o^2\rangle$ are
zero by symmetry.  Therefore, from (\ref{kubocond}), we get
\begin{eqnarray}
\lefteqn{\frac{\sigma_{xy}^H(i\omega_n)}{H}=
\frac{2\pi i}{|\omega_n|\Phi_0} \int_0^{\beta}\!\!d\tau e^{i\omega_n\tau}
\langle j_{{\bf q}=0}^x(\tau)j_{{\bf q}=0}^y(0){\cal A}_o\rangle,}
\nonumber\\ 
\lefteqn{\frac{\Delta\sigma_{xx}(i\omega_n)}{H^2}=
\frac{2\pi^2}{|\omega_n|\Phi_0^2}\int_0^{\beta}\!\!d\tau 
e^{i\omega_n\tau}\times} \nonumber\\
&&[\langle j_{{\bf q}=0}^x(\tau)j_{{\bf q}=0}^x(0)
{\cal A}_o^2\rangle - 
\langle j_{{\bf q}=0}^x(\tau)j_{{\bf q}=0}^x(0)
\rangle\langle{\cal A}_o^2\rangle].
\end{eqnarray} 
Since the oriented area ${\cal A}_o$ can be written as 
\begin{eqnarray}
{\cal A}_o&=&\frac{1}{2}\sum_{\bf r}\int_0^{\beta} \hat{z}\cdot({\bf j}_{\bf 
  r}(\tau)\times{\bf r})\, d\tau\nonumber\\
&=& 
  \frac{1}{2}\int_0^\beta \hat{z}\cdot\left[\partial_{\bf q}\times
  {\bf j}_{\bf q}(\tau)\right]_{{\bf q}=0}\, d\tau \quad , 
\end{eqnarray}
we see that we can related this expression for the Hall conductivity
$\sigma_{xy}^H$ to the more familiar one involving the average of
three currents\cite{fuku}.

Again, we see that the magnetotransport response is strongly
suppressed by the gauge fluctuations because it is sensitive to the
oriented area of the world-line configurations. In principle, the
quantities Im$\sigma_{xy}^H(\omega)$ (from which we can obtain
Re$\sigma^H_{xy}(0)$ from a Kramers-Kronig relation) and
$\Delta\sigma_{xx}(\omega)$ can be computed in a similar way to the
calculation of the optical conductivity. However, these quantities are
too small to measure in the regime of strong gauge fluctuations that
we study.

Since we have argued that the gauge field fluctuations are indeed
strong in the cuprates at temperatures above the superconducting
transition, it appears that our simple boson model with a quasistatic
gauge field cannot describe quantitatively the magnetotransport in
these materials. This result is however qualitatively consistent with
the experimental finding that these magnetotransport properties are
generally suppressed from the classical values. To obtain a
quantitative prediction for these properties, one may attempt to
restore dynamics to the gauge fields. If the gauge field may relax in
time, then the boson world lines no longer have to obey the condition
of strictly retracing paths. This would allow the world lines to
enclose a finite oriented area and hence a finite response to external
magnetic fields. However, we emphasize that such an approach might not
represent the physics completely. We believe that our model
illustrates the general point that the influence of an external field
on the system is strongly masked by the fluctuations of the internal
magnetic field.

\section{density correlation function}

\subsection{Phase separation}

Non-interacting bosons are infinitely compressible. They would
therefore collapse into a small region of the system in the presence
of any quenched disorder which has a tail of localized states in the
single-particle spectrum. An analogous collapse is also found in this
problem with annealed random flux.  Such an instability was discussed
by Feigelman {\em et al.}\cite{feigel} who have argued that it occurs
also in the case of interacting bosons at low densities, leading to a
hole-rich phase and a hole-absent phase. They further argued a
long-range Coulomb repulsion would be necessary to stabilize the
uniform phase.

Within the world-line picture, one can visualize the instability of
the homogeneous phase in the limit of strong gauge fluctuations. The
condition of retracing paths in this limit encourages the bosons to
come close to each other so that their paths may retrace each other.
This will allow individual boson paths to explore a larger area (in
imaginary time), and hence lower the kinetic energy of the system
compared to the case with each boson has to retrace its own path. In
the absence of any repulsive interactions, this effect would dominate
at low temperatures, making the homogeneous phase unstable to
collapse.

We find that this instability towards the formation of dense
aggregates indeed occurs in our model in the absence of boson
repulsion, although the instability is prevented by on-site repulsion,
at least for the moderate boson densities of interest here. We have studied
the instability by examining the compressibility of the
system: $\kappa=\lim_{q\to 0}\kappa({\bf q})$ with
\begin{equation}
  \kappa({\bf q})=\frac{1}{Nn_b}\int_0^{\beta}d\tau
  \langle n_{\bf q}(\tau)n_{\bf -q}(0)\rangle,
\end{equation}
where $n_{\bf q}(\tau)$ is the Fourier transform of the boson
density at imaginary time $\tau$. Alternatively, 
$\kappa =\lim_{q\to 0} \beta S({\bf q})/n_b^2$ where $S({\bf q})$ is the static structure factor: 
\begin{equation}
  S({\bf q}) =\frac{1}{L^2}\langle n_{\bf q}(\tau) n_{-{\bf q}}(\tau)\rangle
\label{strucstatic}
\end{equation}

\begin{figure}[hbt]
\epsfxsize=0.9\columnwidth\epsfbox{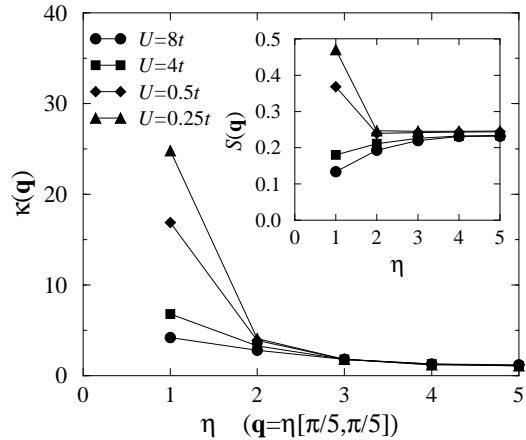} 
\caption{Static structure factor (inset) 
  and the {\bf q}-dependent compressibility as a function of
  $q$ in the ($\pi$,$\pi$) direction for different values of $U$.
  $\beta t=4$, $\fluxvar=0.5\Phi^2_0$, $n_b=0.25$.}
\label{struccomp}
\end{figure}

In Fig.~\ref{struccomp}, we show the behavior of $S({\bf q})$ and
$\kappa({\bf q})$ for different values of the on-site repulsion $U$
for a $10\times 10$ lattice with 25 bosons. The structure factor
$S({\bf q})$ as a function of ${\bf q}$ is qualitatively different for
the cases of small $U$ and large $U$ (compared to $t$): $S({\bf q})$
for ${\bf q}=(\frac{\pi}{5}, \frac{\pi}{5})$ is greater than the
density $n_b$ for when the on-site interaction is small. We can also
look at the compressibility. Since we work with finite systems at
fixed boson number, we will evaluate $\kappa({\bf q})$ at the smallest
wavevector of the system as an estimate of the $q=0$ behavior. We see
that, in the presence of random magnetic flux, the compressibility
increases with decreasing $U$. This can be interpreted as a divergence
as $q\to 0$ for small $U$, and hence an instability of the homogeneous
phase. (This is also reflected in the magnitude of the fluctuations in
our QMC results for $\kappa({\bf q})$ which grows as $q\to 0$ for
sufficiently small $U$.) However, for strong on-site repulsion, the
density correlations show no sign of an instability at this density.

\subsection{Static structure factor}

The density fluctuations in our boson model should be relevant to the
charge fluctuations in the full $t$-$J$ model. It has been pointed out
that the density excitations of the $t$-$J$ model does not resemble those
of a conventional Fermi system. We will now compare our results with
numerical results on the full $t$-$J$ model in the literature.

\begin{figure}[hbt]
\epsfxsize=0.9\columnwidth\epsfbox{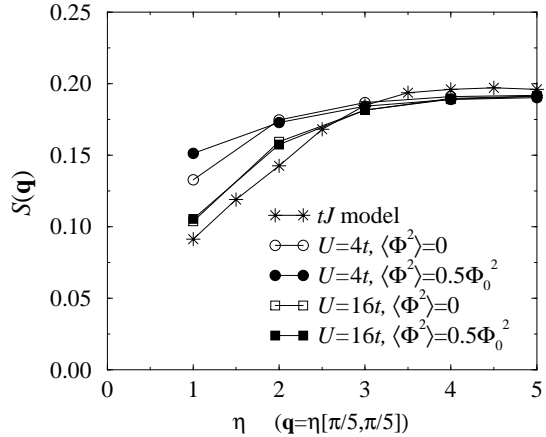} 
\caption{Static structure factor of the boson model at density $n_b=0.2$ 
  along the ($\pi$,$\pi$) direction at $T=0.25t$. Asterisks: $t$-$J$ model
  result\protect\cite{putikka} at electron density $n=1-n_b=0.8$ and
  $t/J=2$.}
\label{sstruc}
\end{figure}

The static structure factor (\ref{strucstatic}) has been calculated by
various means\cite{putikka,ycchen}. Fig.~\ref{sstruc} shows
the static structure factor for our boson system together with that of
the \tJ model\cite{putikka} at $T=0.25t$. We see that our results are
qualitatively similar to the $t$-$J$ model, with improving
quantitative agreement as one approaches the hard-core limit (see, for
example, $U=16t$). We should point out that this dependence on $U$
should not be as strong for the transport properties of the system,
because the particle currents are not directly affected by the
repulsive density interactions.

It is also interesting to note that the magnitude of the gauge field
fluctuations has a relative weak effect on $S({\bf q})$ when the
on-site repulsion $U$ is strong. However, as we shall see
in the next section, the {\em dynamics\/} of the density excitations is
strongly modified by the interaction with the gauge fields.

\subsection{Dynamic structure factor}

We now look at the dynamic structure factor $S({\bf q},\omega)$:
\begin{equation}
  S({\bf q},\omega) = \frac{1}{L^2}\int\!\!dt\, e^{i\omega t}
  \langle n_{\bf q}(t)n_{-{\bf q}}(0)\rangle
\end{equation}
where $n_{\bf q}(t)$ is the Fourier transform of the density in real
time.  
The dynamic structure factor is related to the imaginary-time
density-density correlation function by
\begin{equation}
\frac{1}{L^2}\langle n_{\bf q}(\tau)n_{-\bf q}(0)\rangle\!
=\!\int_0^\infty\!\!\!\!
(e^{-\tau\omega}\!\! +\! e^{-(\beta-\tau)\omega})S({\bf
q},\omega)d\omega.
\label{densME}
\end{equation}
Again, we use MaxEnt to perform the inversion of this integral
equation. Two sum rules can be used as a check of the MaxEnt
procedure.
\begin{eqnarray}
 \int_0^\infty\!\!\! d\omega 
 (1-e^{-\beta\omega})\omega S({\bf q},\omega)&=&
 -\frac{\langle K\rangle}{2L^2}(2-\cos q_x-\cos q_y)t \nonumber\\
 \int_0^\infty\!\!\! d\omega
 \frac{1-e^{-\beta\omega}}{\omega}S({\bf q},\omega)
 &=&\frac{1}{2}n_b^2 \kappa({\bf q})
\end{eqnarray}  
These are lattice versions of the $f$-sum rule and the compressiblity
sum rule. They are satisfied within $1\%$ error in our results.

\begin{figure}[hbt]
\epsfxsize=0.9\columnwidth\epsfbox{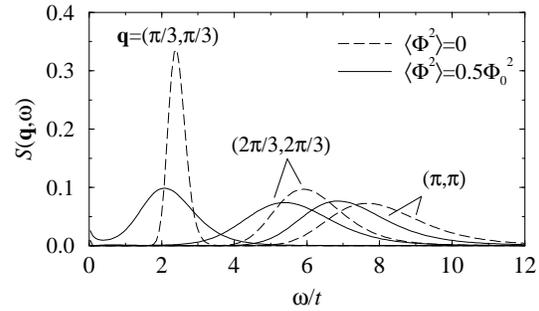} 
\caption{Dynamic structure factor of the superfluid phase
  ($\fluxvar=0$) and the normal phase ($\fluxvar=0.5\Phi^2_0$) in the
  $(\pi,\pi)$ direction. $U=4t$, $T=t/6$ at quarter-filling.}
\label{dystruc}
\end{figure}

Fig.~\ref{dystruc} shows $S({\bf q},\omega)$ for our bosons with and
without the random flux. The system in the absence of random flux
should be a superfluid at the temperature and densities considered
here, and therefore should possess well-defined phonon excitations. We
see sharp phonon peaks in the density excitation spectrum, for
instance, at wavevector ${\bf q}=(\frac{\pi}{3},\frac{\pi}{3})$.
These long-lived phonon excitations of the superfluid phase do not
survive the coherence-breaking effect of the gauge-field interactions.
We find only break peaks in $S({\bf q},\omega)$ in the presence of
strong random flux.

Another effect of the presence of the gauge field is a reduction in
the bandwidth of the density excitations. This might be expected
because the gauge-field interaction tends to increase the
compressibility of the system. Indeed, we see that the center of the
$(\pi,\pi)$ peak is pulled in from $7.6t$ to $6.8t$.

\begin{figure}[hbt]
\epsfxsize=0.9\columnwidth\epsfbox{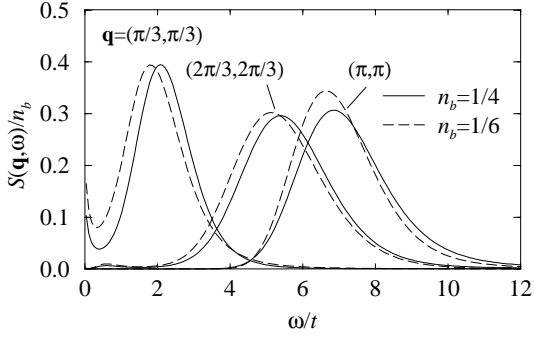} 
\caption{Scaling of $S({\bf q},\omega)$ with boson density in the
  $(\pi,\pi)$ direction.  The solid(dashed) lines are $S({\bf
  q},\omega;n_h)/n_h$ for 9(6) bosons on a 6$\times$6 lattice. $\beta
  t=6$, $U=4t$, $\fluxvar=0.5\Phi^2_0$. }
\label{dscaling}
\end{figure}

We also see that the dynamic structure factor has a simple scaling
with the hole density (Fig.~\ref{dscaling}): $S({\bf
q},\omega; n_h) = n_h S^0({\bf q},\omega)$ holds for $n_h=0.1\sim
0.3$. This is natural in a model of degenerate bosons where the boson
density is equal to the hole density.

We will now compare our results with numerical results on the full
$t$-$J$ model\cite{horsch,eder}. It should be noted that,
although we expect the electron density excitations of the $t$-$J$
model to be dominated by its holon component, there is no quantitative
equivalence between the structure factors of the $t$-$J$ model and our
boson-only model. Nevertheless, we argue that the dynamic structure
factor of our model has qualitative similarities with that of the
$t$-$J$ model. For instance, the absence of sharp peaks in the dynamic
structure factor is also found in the $t$-$J$ model. An obvious
similarity, built into our boson model {\em a priori}, is the lack of
any structure indicating scattering across a Fermi surface at
$q=2k_{\rm F}$. Another feature is the scaling of dynamic structure
factor with the hole density \cite{eder}.

\begin{figure}[hbt]
\epsfxsize=0.9\columnwidth\epsfbox{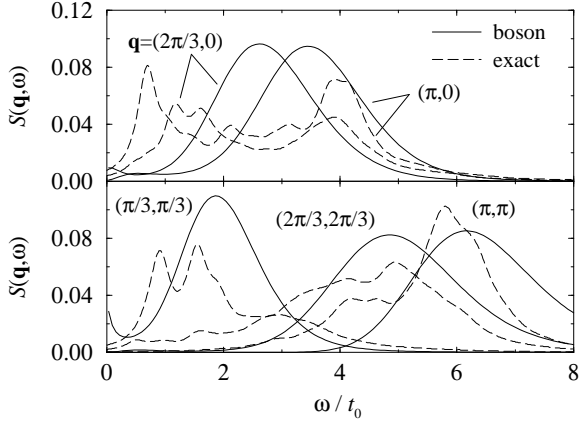} 
\caption{Dynamic structure factor. Solid lines denote our Monte Carlo
  results for 6$\times$6 lattice with 9 bosons at $\beta t=6$ with
  $t=0.9t_0$. $\fluxvar=0.5\Phi_0^2$ and $U=4t$. Dashed
  lines denote exact diagonalization results\protect\cite{eder} for 4
  holes in an 18-site cluster with $t_0/J=2.5$.}
\label{dystruc2}
\end{figure}

We find that $S({\bf q},\omega)$ along the $(\pi,\pi)$ direction
agrees well with an exact diagonalization study of the $t$-$J$ model
at a similar hole density, as shown in Fig.~\ref{dystruc2}. (We have
used a moderate rescaling of the hopping energy: $t=0.9t_0$ where
$t_0$ is the hopping energy in the $t$-$J$ model.) The area under ${\bf
q}=(\frac{\pi}{3},\frac{\pi}{3})$ peak is larger in our model than in
the $t$-$J$ model. We believe that, as in the case of the static
structure factor, this discrepancy can be improved with if we use a
stronger on-site repulsion. However, the structure factor does not
agree with the $t$-$J$ model along the $(\pi,0)$ direction. It might
be that the spectrum of the holes at zero temperature is qualitatively
different from the simple tight-binding spectrum that we have assumed
here.

\section{conclusion}

In summary, we have studied a degenerate Bose system which remains
metallic below its degeneracy temperature due to elastic scattering
with random and quasistatic gauge fields. In the path-integral
picture, the bosons retrace their paths in the limit of strong gauge
fluctuations in order to avoid the quantum frustration due to the
fluctuating gauge field. We have demonstrated that many features of
these ``Brinkman-Rice bosons'' indeed mimic the behavior of the full
$t$-$J$\ model and the normal state of the cuprate superconductors.
These features include the linear-$T$ dependence of the longitudinal
scattering rate and a charge excitation spectrum which consists of
broad incoherent structures. This model itself has a strongly
suppressed response to external magnetic fields, hinting that the
behavior of the system as measured in Hall and magnetoresistance
experiments have to understood in terms of a separate mechanism.

It would also be interesting to understand the behavior of the system
in the zero-temperature limit. Although the limit of infinite gauge
fluctuations ({\em i.e.}, a uniform flux distribution on a lattice)
would strictly forbid any world lines to wrap around periodic
boundaries, one may consider the case of weaker gauge fluctuations in
the zero-temperature limit and ask whether there is a critical value
of $\fluxvar$ below which the system is a superfluid at zero
temperature. This will involve a study of the system at very low
temperatures near a quantum critical point. This is beyond the scope
of this paper.

We thank Wolfgang von der Linden for sending us his MaxEnt code and
for many helpful correspondences.  We also thank W. Putikka, R. Eder,
and S. Maekawa for sending us their data for comparison. We
acknowledge helpful conversations with J.T.~Chalker, S.M.~Girvin, D.H.~Lee,
E.~Sorensen, X.-G.~Wen, and S.C.~Zhang. This work was supported by the
NSF MRSEC program (DMR 94-0034), NEC (DHK) and EPSRC/NATO (DKKL).

\begin{appendix}
\section{Operator Averages}

In this appendix, we discuss the evaluation of operator averages.  In
the path-integral representation of the partition function, world-line
configurations are sampled according to a distribution proportional to
$\exp(-S_B^0-S_2)$. In our Monte Carlo scheme, we discretize the
imaginary-time interval $\beta$ into $M$ segments of length
$\Delta\tau$, and we define a world-line configuration by the boson
coordinates at these discrete time points:
$\{R_0,R_1,R_2,\ldots,R_M=P(R_0)\}$. ($R_m$ denotes the coordinates of
the $N$ bosons at $m$-th time slice: $R_m=({\bf x}_1^{(m)},\ldots,{\bf x}_N^{(m)})$, and $R_M$ is a permutation of the coordinates of
$R_0$.)

The configurations are sampled according to the probability:
\begin{equation}
{\cal P}(\{R\})=\frac{1}{\cal N}
e^{-S_2(\{R\})}\prod_{m=0}^{M-1}\rho_{\Delta\tau}(R_m,R_{m+1}),
\label{sample}
\end{equation}
where ${\cal N}$ is a normalization constant, and
$\rho_{\Delta\tau}(R,R')$ is the short-time (high-temperature) density
matrix in the absence of gauge fields. It is given by:
\begin{eqnarray}
\rho_{\Delta\tau}(R,R')&=&\langle R|e^{-\Delta\tau (H_K^0+H_U)}|R'\rangle\nonumber\\
&\simeq&\langle R|e^{-\frac{1}{2}\Delta\tau H_U}e^{-\Delta\tau H_K^0}
 e^{-\frac{1}{2}\Delta\tau H_U}|R'\rangle\nonumber\\
&=&\langle R|e^{-\Delta\tau H_K^0}|R'\rangle 
 e^{-\frac{1}{2}\Delta\tau (\tilde{H}_U(R)+\tilde{H}_U(R'))}
\label{potential}
\end{eqnarray}
with $\tilde{H}_U(R)=\langle R| H_U | R \rangle$, and
\begin{eqnarray}
H_K^0 & =&  -t\sum_{\langle ij \rangle}
(b_{i}^{\dagger}b_{j} +{\rm h.c.}) \\
H_U &=& \frac{U}{2} \sum_i n_i(n_i-1).
\end{eqnarray}
The error involved in this approximation of the density matrix is
O$(\Delta\tau^3)$. 

Measurements which depend only on particle positions are simple to
evaluate in this path-integral representation. The expectation value of such a measurement ${\cal O}$ is given by
\begin{equation}
\langle{\cal O}\rangle={\rm Tr}\left[\tilde{{\cal O}}(R_0,R_1,R_2,\ldots )
{\cal P}(R_0,R_1,R_2,\ldots )\right]
\end{equation}
where $\tilde{{\cal O}}(\{R\})$ is the measured value for the
world-line configuration $\{R\}$, and the trace is taken over all
such configurations.

Averages of operators which are non-local in position space are more
cumbersome to evaluate. An example is the kinetic energy:
\begin{equation}
\langle H_K \rangle = -t\sum_{\langle ij \rangle}
\langle (e^{i a_{ij}} b_{i}^{\dagger}b_{j} +{\rm h.c.}) \rangle.
\label{kinetic}
\end{equation}
The gauge field $a_{ij}$ is defined on the link between the
neighboring sites $i$ and $j$. The Peierls factor closes the gap in
the imaginary-time loop caused by the action of kinetic energy
operator. Inserting the operator $H_K$ in the imaginary-time slice
between the $m=0$ and $1$, it can be shown that the kinetic energy can
be evaluated as:
\begin{eqnarray}
\langle H_K \rangle&=& {\rm Tr} \left[\frac{\langle R_0|H_K^0
{\cal U}_{\Delta\tau}|R_1\rangle}
{\langle R_0|{\cal U}_{\Delta\tau}|R_1\rangle}
{\cal P}(\{R\})\right]\nonumber\\
&=& {\rm Tr}\Bigg[\frac{\langle R_0| H_K^0
e^{-\Delta\tau H_K^0}|R_1\rangle}
{\langle R_0|e^{-\Delta\tau H_K^0}|R_1\rangle}\times\nonumber\\
&&\qquad
e^{\frac{1}{2}\Delta\tau
(\tilde{H}_U(R_0)-\tilde{H}_U(KR_0))}{\cal P}(\{R\})
\Bigg]
\end{eqnarray}
where $\tilde{H}_U(KR)$ is defined by $H_U \{H_K^0|R\rangle \}\equiv
\tilde{H}_U(KR) \{H_K^0|R\rangle \}$, and ${\cal U}_{\Delta\tau}$ is
the short-time evolution operator:
\begin{equation}
{\cal
U}_{\Delta\tau}=e^{-\frac{1}{2}\Delta\tau H_U} e^{-\Delta\tau H_K^0}
e^{-\frac{1}{2}\Delta\tau H_U}.
\end{equation}
\end{appendix}

\begin{table}[hbt]
\caption{One, two, three, and four- boson exchange probability for various
 $T,\fluxvar$, and $U$ at quarter-filling.} 
\label{tab1}
\begin{tabular}{lllllll} 
$T$ & $U$ & $\fluxvar/\Phi_0^2$ & ${\rm P}_1$ & ${\rm P}_2$ & $ {\rm P}_3$ & $ {\rm P}_4$ \\ \hline
\tableline 
0.5$t$ & 4$t$ & 0.5 & 0.51 & 0.23 & 0.13 & 0.07\\
0.5$t$ & 4$t$ & 0 & 0.20 & 0.12 & 0.11 &0.11 \\ \hline
0.25$t$ & 4$t$ & 0 & 0.12 & 0.11 & 0.11 & 0.11 \\
0.25$t$ & 4$t$ & 0.5 & 0.26 & 0.16 & 0.13 & 0.12\\
0.25$t$ & 16$t$ & 0.5 & 0.41 & 0.21 & 0.13 & 0.10\\  \hline
0.11$t$ & 4$t$ & 0 & 0.11 & 0.11 & 0.11 & 0.11\\
0.11$t$ & 4$t$ & 0.5 & 0.12 & 0.11 & 0.11 & 0.11\\ 
0.11$t$ & 16$t$ & 0.5 & 0.12 & 0.11 & 0.11 & 0.11\\ 
\end{tabular} 
\end{table}

\end{multicols}

\begin{references}
\bibitem{gauss} However, high temperature series studies of the gauge
  field model\cite{hlubina} have found that at least the thermodynamic
  properties of the $t$-$J$ model are well described by the gauge
  field model within the Gaussian approximation.
\bibitem{wheatcoh} In Ref.\ \onlinecite{wheathong}, Wheatley and Hong have
 defined a ``coherence temperature'' $T_{\rm coh}$ in a similar way as
 our $T_D$.  In their case of bosons coupled to dissipative bath, it
 was argued that $T_{\rm coh}$ is pushed down to zero.
\end{references}
\end{document}